\documentclass[letterpaper,journal]{IEEEtran}

\usepackage[utf8]{inputenc}
\usepackage{algorithm}
\usepackage{algpseudocode}
\usepackage{amsmath} % for \DeclareMathOperator
\usepackage{graphicx}
\usepackage{caption}
\usepackage{subcaption}
\usepackage{booktabs}
\usepackage{graphicx} % Required for \resizebox
\usepackage{multirow}

\usepackage{amsbsy,amsmath,amssymb,epsfig,bbm,mathrsfs,multirow,amsthm}
\usepackage{float}
\usepackage{cite}
\usepackage{array, multirow, graphicx}
\usepackage{graphics}
\usepackage{makecell}
\usepackage{comment}
\usepackage{algpseudocode} % Use either algorithmic or algorithmicx, not both
\usepackage{hyperref}
\usepackage{color}
\usepackage{lipsum}
\usepackage{caption}
\usepackage{subcaption}
\usepackage{multirow}
\usepackage{svg}
\usepackage{placeins}

\newcommand{\changes}[1]{\textcolor{black}{#1}}

\begin{document}

%\title{Efficient Deployment of Domain-Specific Multimodality Foundation Models via Federated Fine-Tuning in Resource-Constrained Edge Networks}

\title{CARE-FL: Federated Fine-Tuning of Multimodal LLMs in Resource-Constrained Edge Networks}

\author{Tianqi~Zong, Baosheng~Li, Nuocheng~Yang,~\IEEEmembership{Student~Member,~IEEE}, \\Binbin~Chen,~\IEEEmembership{Senior~Member,~IEEE}, Marie~Siew, ~\IEEEmembership{Member,~IEEE}

\thanks{Tianqi~Zong, Binbin~Chen, and Marie~Siew are with the Information Systems Technology and Design, Singapore University of Technology and Design, 487372, Singapore (emails: zong\_tianqi@mymail.sutd.edu.sg, binbin\_chen@sutd.edu.sg ,marie\_siew@sutd.edu.sg).}

\thanks{Baosheng Li is with the College of Science, Northwest A\&F University,
Yangling 712100, China (email: baoshengli96@nwafu.edu.cn).}

\thanks{Nuocheng Yang is with the Beijing Laboratory of Advanced Information Network, and the Beijing Key Laboratory of Network System Architecture and Convergence, Beijing University of Posts and Telecommunications, Beijing 100876, China (email: yangnuocheng@bupt.edu.cn).}

}

\maketitle

\begin{abstract}
Domain-specific adaptation of Multimodal Large Language Models (MLLMs) in privacy-sensitive edge environments requires federated fine-tuning, yet client selection remains challenging due to heterogeneous multimodal data and modality availability, dynamic system conditions, and evolving client utility during training. We propose Constraint-Aware Reinforcement Learning for Efficient Federated Fine-Tuning (CARE-FL), a constrained online client-selection framework for federated MLLM Low-Rank Adaptation (LoRA) fine-tuning. CARE-FL formulates client participation as a sequential decision process and employs Q-value decomposition to address the large combinatorial joint action space through scalable client-level long-term utility estimation. These utilities are learned online using a modified Double Deep Q-Learning (DDQL) scheme, while constraint-aware \(\epsilon\)-greedy execution selects feasible client subsets under per-round latency and bandwidth constraints. Complementing client selection,  an adaptive aggregation mechanism further controls potentially unstable global updates and \changes{places greater weight on the previous model when needed.} %retains the previous model when recovery attempts fail% 
Experiments on the CrisisMMD disaster-response task in emulated heterogeneous wireless edge environments show that CARE-FL achieves higher and more stable test accuracy than the compared client-selection baselines across different system settings, while satisfying the imposed latency and bandwidth constraints.
\end{abstract}

\begin{IEEEkeywords}
Federated learning, multimodal LLMs, edge intelligence, reinforcement learning, client selection.
\end{IEEEkeywords}

\section{Introduction}
% paragraph1:
Recent advances in Large Language Models (LLMs) and Multimodal Large Language Models (MLLMs) have significantly improved the capabilities of deep learning systems in complex understanding, reasoning, and generation tasks \cite{vaswani2017attention, zhang2024mm}. By jointly processing multimodal information sources such as text, images, audio, and sensor data, MLLMs better reflect the multimodal nature of real-world human interactions and have demonstrated promising performance across diverse applications, such as healthcare \cite{alsaad2024multimodal}, finance \cite{liu2025multimodal}, education \cite{borazjani2025bringing}, biomedical analysis \cite{li2025open}, and transportation \cite{xu2024drivegpt4}. Consequently, domain-specific fine-tuning of MLLMs is becoming increasingly important for deploying intelligent services in practical scenarios.

% Paragtaph2:
%However, centralized fine-tuning of well-generalized domain-specific MLLMs remains challenging because it requires large amounts of domain-specific data distributed across multiple sources. 
\changes{However, centralized fine-tuning of MLLMs requires collecting task-relevant data from distributed sources, which is often impractical in privacy-sensitive scenarios.} \changes{ Federated Learning (FL) enables edge clients to collaboratively train a global model by exchanging model parameters or updates without sharing raw data \cite{mcmahan2017communication, siew2025fair}.}
%Federated Learning (FL) provides a promising framework that enables distributed edge nodes to collaboratively fine-tune a shared global model without transmitting raw local data \cite{mcmahan2017communication}. 
Moreover, MLLMs are increasingly supported on common mobile and edge platforms, including iOS, Android, and HarmonyOS, with several edge-adaptation frameworks and deployment toolchains becoming available \cite{yao2024minicpm}.  Together with Parameter-Efficient Fine-Tuning (PEFT) techniques such as Low-Rank Adaptation (LoRA) \cite{hu2021lora}, which reduces the number of trainable parameters and communication overhead of large-model adaptation, these developments make federated MLLM fine-tuning increasingly feasible in resource-constrained edge environments. 
For example, recent studies have explored federated PEFT of LLMs for the Internet of Vehicles \cite{xie2024gai} and distributed disaster response \cite{zhao2025energy}.

Despite the increasing feasibility of federated PEFT of MLLMs on edge devices, practical deployment in wireless edge environments remains challenging. %particularly for latency-sensitive applications such as autonomous driving and disaster-response coordination. 
On the learning side, clients exhibit statistical heterogeneity in data quantity, distribution, and modality composition due to differences in sensing capabilities, device types, user behaviors, and observation environments, resulting in varying contributions from clients to global model learning. On the system side, edge clients exhibit heterogeneous and resource-limited computational and communication capabilities \changes{\cite{xie2024gai, zhao2025energy}. For example, autonomous vehicles and UAVs may have limited onboard processing capacity and wireless bandwidth.} Device workloads and wireless channel conditions also vary over time. Together with limited shared network resources, these factors create heterogeneous and time-varying participation costs, further complicating federated PEFT of
MLLMs in resource-constrained edge environments.

Existing studies have improved the efficiency of federated PEFT of MLLMs in wireless edge environments through model-side optimization strategies, such as PEFT trainable-parameter configuration optimization \cite{liu2025adaptive, fang2025automated, liu2025fedquad}, as well as through joint model adaptation and wireless resource management \cite{zou2025joint, wu2025resource, zhang2025communication}. These approaches primarily focus on model configuration and communication-resource optimization; however, client participation often relies on uniform random sampling, leaving client-selection optimization relatively underexplored. \changes{Moreover, limited communication resources and the high computational demands of MLLM fine-tuning restrict the number of training rounds achievable within a given time budget, highlighting the importance of effective client selection alongside resource optimization.}
%\changes{ Moreover, limited communication resources and the high computational demands of MLLM fine-tuning may restrict limited client participation in limited total training time or number of training rounds available to achieve satisfactory model performance, making effective client selection becomes explicitly important.}

Extensive studies in traditional federated learning have demonstrated that client participation significantly affects model convergence and system efficiency under statistical and system heterogeneity. Consequently, numerous client-selection strategies have been proposed to prioritize participants according to their estimated training utility, resource efficiency, or both \cite{fu2023client}\changes{, where training utility represents a client's potential contribution to global model learning.} However, directly extending these approaches to federated MLLM fine-tuning is non-trivial due to complex multimodal semantic data and substantially higher computation and communication costs. These characteristics introduce new challenges.

%Compared with conventional FL tasks, MLLM fine-tuning involves richer semantic content and heterogeneous multimodal inputs, while large-model adaptation incurs substantially higher computation and communication costs. 

%\changes{MLLM fine-tuning operates on semantically complex multimodal data, such as images and natural-language text, while large-model adaptation incurs substantially higher computation and communication costs. These characteristics introduce new challenges in estimating client training utility and ensuring system-efficient participation in heterogeneous edge environments, as discussed below.}
%\textcolor{red}{Nuocheng: We need to clarify the definition of "richer semantic content" of MLLM fine-tuning?}

Firstly, client training utility is more difficult to characterize in federated MLLM fine-tuning due to complex semantic and multimodal statistical heterogeneity. Beyond differences in local data quantity and distribution, clients may possess different modality combinations, while the semantic content of multimodal data can further lead to heterogeneous local training behaviors. Recent studies \cite{zhang2025data} have explored LLM-based data-aware client selection using knowledge-graph representations of local datasets.  Although such approaches characterize semantic data divergence, they may introduce privacy risks, face difficulties with heterogeneous multimodal inputs, and rely on relatively static client representations that do not adapt to evolving fine-tuning states.

%rely on relatively static client representations. In federated MLLM fine-tuning, however, the training utility of a client may also vary with the evolving global model state, making static utility characterization insufficient for adaptive participation decisions.

Secondly, system-efficient client participation is more difficult to achieve under heterogeneous and dynamic edge conditions. Different local data volumes and modality combinations lead to heterogeneous computational workloads, while device capabilities and wireless conditions introduce further variability in participation costs. Recent studies \cite{zhao2025movable} have addressed system efficiency through joint client selection and wireless resource allocation under emerging technologies such as movable antennas and over-the-air aggregation. However, these approaches primarily focus on system-level resource optimization, without explicitly incorporating client training utility under statistical and modality heterogeneity into participation decisions.

These limitations highlight the need for adaptive client selection that jointly considers evolving training utility and dynamic system feasibility. Moreover, client selection is inherently sequential: current participation decisions influence subsequent global model states and future client utilities. These characteristics highlight the need for an adaptive online client-selection strategy that jointly considers evolving client training utility, dynamic system feasibility, and the long-term impact of participation decisions.

In view of the above limitations, we propose Constraint-Aware Reinforcement Learning for Efficient Federated Fine-Tuning (CARE-FL), an online client-selection framework for federated PEFT of MLLMs in resource-constrained wireless edge environments. The main contributions are as follows:

\begin{itemize}
\item We propose Constraint-Aware Reinforcement Learning for Efficient Federated Fine-Tuning (CARE-FL), an end-to-end framework for federated PEFT of MLLMs in resource-constrained wireless edge environments. CARE-FL integrates LoRA-based fine-tuning, RL-assisted dynamic client selection, and adaptive aggregation under heterogeneous multimodal data and system conditions.

\item %We formulate adaptive client participation as a constrained sequential decision-making problem and establish an MDP framework that captures evolving client-specific model states reflecting heterogeneous data and modality characteristics, as well as the long-term impact of participation decisions. Per-round latency and bandwidth constraints are incorporated to ensure system feasibility.
We formulate adaptive client participation as a constrained sequential decision-making problem and establish an MDP framework \changes{that captures evolving client-specific states through compact PCA-based model representations together with heterogeneous data and modality characteristics, as well as the long-term impact of participation decisions. Per-round latency and bandwidth constraints are incorporated to ensure system feasibility.}

\item We develop a constraint-aware Q-value-decomposed DDQL scheme for scalable online client selection. \changes{ Q-value decomposition replaces the representation of \(\binom{K}{n}\) subset-level Q-values with \(K\) client-level long-term utility estimates, substantially reducing the Q-value representation and search complexity,} while modified DDQL continuously learns these utilities from evolving fine-tuning feedback. Constraint-aware \(\epsilon\)-greedy execution enables feasible exploration and exploitation under the system constraints. Complementing client selection, an adaptive aggregation mechanism mitigates potentially unstable global updates.

\item Extensive experiments \changes{on the CrisisMMD dataset demonstrate that CARE-FL achieves 3--5 percentage points} higher test accuracy with greater stability than the compared client-selection baselines while satisfying the imposed latency and bandwidth constraints, and remains effective across different PEFT configurations and wireless resource conditions.
\end{itemize}

%The remainder of this paper is organized as follows. Section II reviews the related work. Section III presents the system model for federated MLLM fine-tuning in heterogeneous wireless edge environments. Section IV formulates the client-selection optimization problem and discusses the associated challenges. Section V presents the proposed CARE-FL framework, including RL-assisted client selection and adaptive aggregation. Section VI presents the experimental evaluation, and Section VII concludes the paper.

The remainder of this paper is organized as follows. Sections II–IV present the related work, system model, and problem formulation, respectively. Section V introduces the proposed CARE-FL, Section VI presents the experimental results, and Section VII concludes the paper.

\section{Related Work}

\subsection{LLMs, MLLMs, and Parameter-Efficient Fine-Tuning}
Large Language Models (LLMs), primarily based on the Transformer architecture, have demonstrated strong capabilities in language understanding, reasoning, and generation \cite{vaswani2017attention}. Multimodal Large Language Models (MLLMs) extend these capabilities to modalities such as images, audio, and video. Modern MLLMs commonly employ a modular architecture consisting of modality-specific encoders for feature extraction, lightweight connectors for alignment with the language embedding space, and a pretrained LLM backbone for multimodal understanding, reasoning, and generation \cite{wadekar2024evolution}. Representative MLLMs include Flamingo \cite{alayrac2022flamingo}, BLIP-2 \cite{li2023blip}, LLaVA \cite{liu2023visual}, and MiniCPM-V \cite{yao2024minicpm}.

Due to the large parameter scale of LLMs and MLLMs, full-parameter fine-tuning incurs substantial computation and memory costs. Parameter-Efficient Fine-Tuning (PEFT) addresses this challenge by updating only a small subset of trainable parameters while keeping most pretrained parameters frozen \cite{he2021towards}. Representative approaches include Adapter Tuning \cite{houlsby2019parameter}, Prefix Tuning \cite{li2021prefix}, and Low-Rank Adaptation (LoRA) \cite{hu2021lora}. In particular, LoRA represents model updates using trainable low-rank matrices, substantially reducing the number of trainable parameters while maintaining competitive adaptation performance. Its configurable rank and insertion locations enable flexible trade-offs between adaptation capability and training overhead, making LoRA particularly suitable for resource-efficient federated fine-tuning of LLMs and MLLMs.

\subsection{Federated Learning for LLMs and MLLMs}
%Federated Learning (FL) enables distributed clients to collaboratively train a shared model without directly exchanging their raw local data. 

Combined with PEFT, FL enables federated LLM adaptation by exchanging only lightweight trainable parameters instead of full model weights, reducing communication and computation overhead \cite{zhang2023fedpetuning}. This allows distributed data to be incorporated into pretrained LLMs without centralizing it.

%zheng2025fedvlmbench
More recently, federated PEFT has been extended from text-only LLMs to MLLMs \cite{zhou2024empirical}. Compared with federated LLMs, federated MLLMs introduce additional heterogeneity arising from multimodal data, including different modality combinations, missing modalities, and cross-client differences in multimodal data distributions \cite{xu2024fedmllm}. Such heterogeneity increases the complexity of local data distributions and leads to varying training contributions among clients.

Existing studies on federated large-model PEFT mainly focus on text-based LLMs, exploring parameter-efficient adaptation and communication-resource optimization \cite{liu2025adaptive, fang2025automated, zou2025joint, wu2025resource}. Recent work has begun to extend PEFT to MLLMs. For example, one-shot federated low-rank adaptation \cite{xu2026you} investigates LoRA-based MLLM fine-tuning under a one-shot communication setting. Nevertheless, adaptive client participation under multimodal heterogeneity and dynamic edge constraints, including bandwidth and latency limitations, remains insufficiently investigated.

\subsection{Client Selection in FL}
Due to communication and computational resource constraints, only a subset of clients can participate in each FL round. Consequently, client selection plays an important role in determining both model convergence and training efficiency. Existing client-selection approaches can be broadly categorized into statistical heterogeneity-aware, system heterogeneity-aware, joint heterogeneity-aware, and learning-based adaptive approaches.

Statistical heterogeneity-aware approaches prioritize clients using data quantity and quality, training loss, gradient information, or estimated contributions to global model improvement\cite{cho2022towards, li2022pyramidfl,zhao2022participant}. By identifying clients with higher training utility, these approaches aim to accelerate convergence and improve model performance under non-independent and non-identically distributed (non-IID) data. In contrast, system heterogeneity-aware approaches consider clients' computational and communication capabilities to prioritize resource-efficient clients and mitigate stragglers \cite{chai2020tifl, shin2022fedbalancer}. Joint heterogeneity-aware approaches combine statistical utility and system efficiency to balance convergence performance and training costs \cite{lai2021oort, tang2025joint}. 

Beyond the above selection strategies, learning-based methods, particularly reinforcement learning (RL), learn adaptive client-selection policies from historical training and system information \cite{xu2020client}. By modeling selection as sequential decision-making, RL adapts participation to evolving model and system states rather than relying solely on predefined metrics \cite{wang2020optimizing, bouaziz2023flash, moon2024client}. This is particularly relevant to dynamic wireless edge environments with time-varying client utilities and resource conditions.

%Nevertheless, most existing client-selection methods, including RL-based approaches, target conventional FL and characterize clients mainly through traditional statistical and system information. Federated MLLM fine-tuning introduces additional client heterogeneity in modality availability, modality combinations, and semantic characteristics of local data. Moreover, client utilities evolve with the global model, while system conditions vary across clients and rounds. Therefore, adaptive client selection that jointly accounts for multimodal statistical characteristics, evolving model information, and dynamic system conditions remains insufficiently explored in federated MLLM fine-tuning.
Nevertheless, most existing client-selection methods, including RL-based approaches, focus on conventional FL and rely mainly on traditional statistical and system information. Federated MLLM fine-tuning introduces additional heterogeneity in modality availability, combinations, and local data semantics. Moreover, client utilities evolve with the global model, while system conditions vary across rounds. Therefore, adaptive client selection jointly considering multimodal data characteristics, evolving model information, and dynamic system conditions remains insufficiently explored in federated MLLM fine-tuning.

%In federated MLLM fine-tuning, client contributions are additionally influenced by multimodal data characteristics, including modality availability, modality combinations, and semantic differences across local datasets. Meanwhile, system conditions may vary across clients and communication rounds, and the relative utility of individual clients may evolve with the global model state throughout fine-tuning. Therefore, jointly incorporating multimodal statistical characteristics, evolving model information, and dynamic system conditions into adaptive client participation remains insufficiently explored in federated MLLM fine-tuning.

\begin{figure*}[!t]
	\centering
	\includegraphics[
	    width=0.95\linewidth,
	    trim={0 0 0 3cm},
	    clip
	]{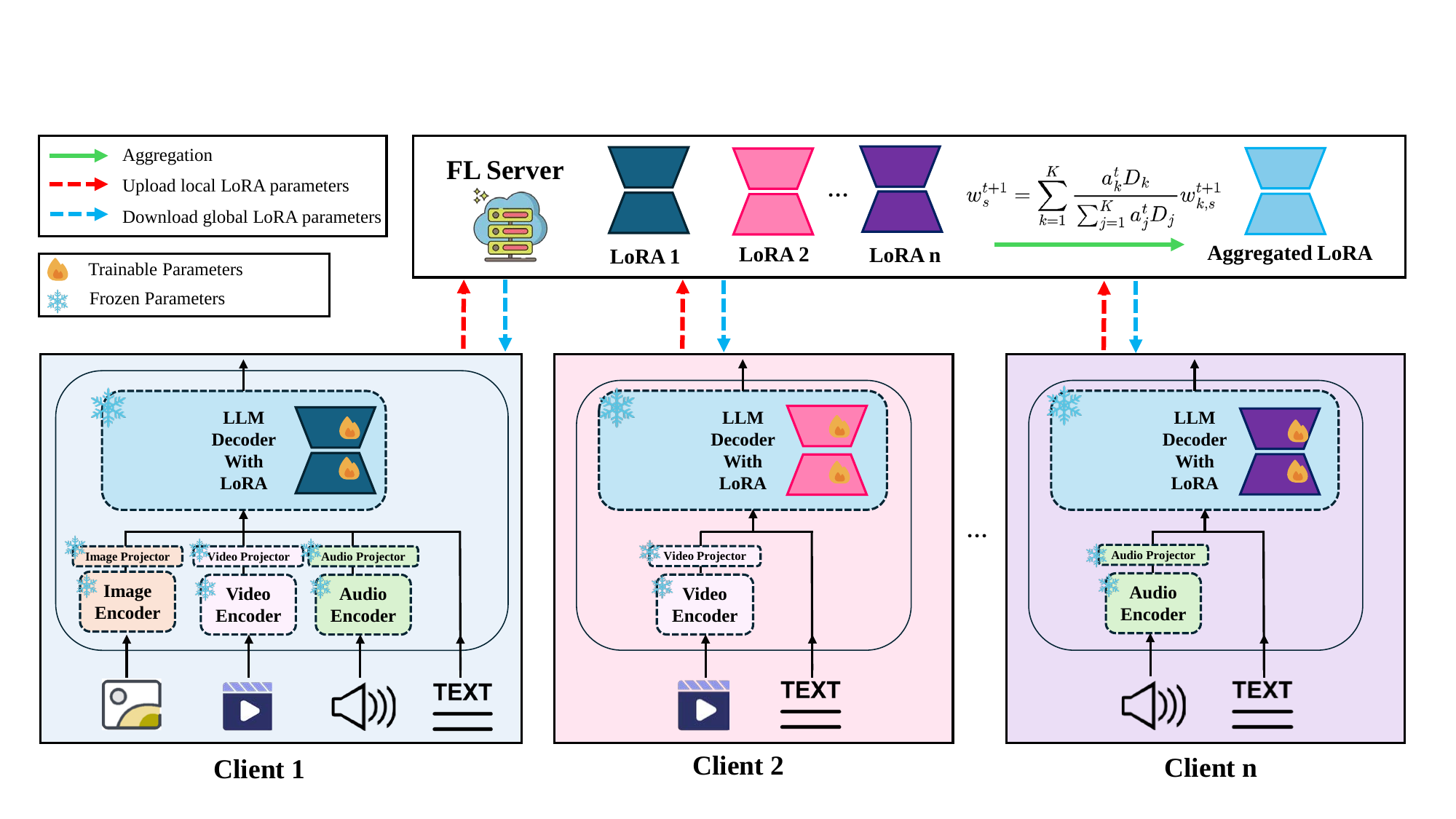}
	\caption{The proposed federated MLLM LoRA fine-tuning framework, where heterogeneous clients process different modalities and the server aggregates LoRA updates while keeping the LLM backbone frozen.}
	\label{fig:framework}
\end{figure*}

%The proposed federated MLLM LoRA fine-tuning framework. The server aggregates lightweight LoRA updates from heterogeneous clients. Each client possesses a distinct modality combination (image, video, and audio) processed by the corresponding modality-specific encoders and projectors, while the LLM backbone remains frozen for communication- and parameter-efficient federated fine-tuning.

\section{System Model}
\subsection{MEC Network Architecture}
We consider a mobile edge computing (MEC) network consisting of a central server and \(K\) edge clients, indexed by \(\mathcal{K}=\{1,\ldots,K\}\). Client \(k\) holds a local dataset
$$ \mathcal{D}_k=\{(X_{k,i},y_{k,i})\}_{i=1}^{D_k}, $$ 
where \(X_{k,i}\) denotes the multimodal input and \(y_{k,i}\) is the corresponding target response sequence. The global dataset is denoted by \(\mathcal{D}=\bigcup_{k\in\mathcal K}\mathcal D_k\), with \(D=\sum_{k=1}^{K}D_k\). Let \(\mathcal M\) denote the set of modalities available in the network and \(\mathcal M_k\subseteq\mathcal M\) the modality subset possessed by client \(k\). Accordingly, \(X_{k,i}=\{x_{k,i}^{m}\}_{m\in\mathcal M_k}\).

%The clients exhibit both statistical and system heterogeneity. Statistical heterogeneity arises from differences in local data volume \(D_k\), label distributions, and modality combinations \(\mathcal M_k\). Meanwhile, system heterogeneity arises from differences in clients' computing and communication capabilities and wireless conditions, which may further vary across training rounds due to dynamic device workloads and wireless channels. We assume that the central server has sufficient computation and communication resources, while the edge clients are resource-constrained. Consequently, different clients exhibit heterogeneous training behaviors as well as computation and communication latency. In the following, we first describe federated MLLM fine-tuning over heterogeneous local datasets and then model the corresponding latency under statistical and system heterogeneity.
Clients exhibit statistical heterogeneity in data volumes, label distributions, and modality combinations, as well as system heterogeneity in computational capabilities and time-varying wireless conditions. We assume a resource-rich central server and resource-constrained edge clients, resulting in heterogeneous training behavior and computation and communication latency. We next describe federated MLLM fine-tuning and its associated latency model.

\subsection{Federated MLLM LoRA Fine-Tuning}
As illustrated in Fig. 1, clients collaboratively fine-tune a pretrained MLLM for a domain-specific task through FL. The global training objective is
\begin{equation} \min_{w_s\in\mathbb{R}^{d}} F(w_s) = \frac{1}{D} \sum_{k=1}^{K}D_k F_k(w_f,w_s),\end{equation} 
where \(F_k(\cdot)\) is the local loss of client \(k\), \(w_f\) denotes the frozen pretrained parameters, and \(w_s\) denotes the trainable parameters shared through FL.
The pretrained MLLM comprises modality-specific encoders and projectors followed by an LLM decoder. Each client uses the encoders and projectors corresponding to its available modalities \(\mathcal M_k\). To reduce local training and communication overhead, LoRA is applied to the LLM decoder while the pretrained backbone remains frozen.  This enables modality-specific feature processing while collaboratively adapting the LLM decoder's reasoning capability. The trainable parameters are
$$ w_s=\{B^{\mathrm{llm}},A^{\mathrm{llm}}\}, $$ 
where \(B^{\mathrm{llm}}\) and \(A^{\mathrm{llm}}\) denote the collections of trainable low-rank LoRA matrices in the LLM decoder. Accordingly, the forward propagation of a LoRA-adapted linear layer in the LLM decoder is
\[
\mathbf{y}
=
\left(\mathbf{W}^{\mathrm{llm}}+\mathbf{B}\mathbf{A}\right)\mathbf{x}
=
\mathbf{W}^{\mathrm{llm}}\mathbf{x}
+\mathbf{B}\mathbf{A}\mathbf{x},\]
where \(\mathbf W^{\mathrm{llm}}\) denotes the frozen pretrained weight matrix of the corresponding linear layer, and \(\mathbf B\) and \(\mathbf A\) denote its trainable LoRA matrices.

\begin{comment}
and the forward propagation of a LoRA-adapted linear layer is
$$\mathbf{y}=\mathbf{W}_{\mathrm{llm}}\mathbf{x}+\mathbf{B}\mathbf{A}\mathbf{x}.$$
\end{comment}
The federated fine-tuning procedure consists of four stages.

\textbf{Stage 1: Initialization.} The server initializes the pretrained MLLM and deploys the model components required by each client's modality combination.

\textbf{Stage 2: Client Selection and Global Model Distribution.} At each round $t \in \{1,\ldots,T\}$,  the server selects \(n\) clients using binary indicators \(a_k^t\in\{0,1\}\), where \(a_k^t=1\) indicates that client \(k\) is selected. The participating set is $$ \mathcal A_t=\{k\in\mathcal K\mid a_k^t=1\}, \sum_{k=1}^{K}a_k^t=n. $$ The server distributes the global LoRA parameters \(w_s^t\) to the selected clients.

\textbf{Stage 3: Local Training and Model Upload.} Each selected client fine-tunes its LoRA parameters using the local autoregressive next-token cross-entropy loss
\begin{equation}
\begin{aligned}
F_k(w_f,w_{k,s})
= {} & -\frac{1}{D_k}
\sum_{i=1}^{D_k}
\sum_{\tau=1}^{|y_{k,i}|} \\
& \log p_{w_f,w_{k,s}}
\left(
y_{k,i,\tau}
\mid y_{k,i,<\tau}, X_{k,i}
\right).
\end{aligned}
\end{equation} 
Each target token is predicted from the multimodal input and preceding target tokens. The backbone \(w_f\) remains frozen and only the local LoRA parameters \(w_{k,s}\) are updated. Padding and other non-target tokens are masked from the loss computation. After local training, each participating client uploads its updated LoRA parameters to the server.

\textbf{Stage 4: Model Aggregation and Global Model Update.} The server aggregates the uploaded LoRA parameters using Federated Averaging (FedAvg):
\begin{equation}
w_s^{t+1} = \sum_{k=1}^{K} \frac{a_k^tD_k}{\sum_{j=1}^{K}a_j^tD_j} w_{k,s}^{t+1}.
\end{equation} 
Since \(w_{k,s}=\{B_k^{\mathrm{llm}},A_k^{\mathrm{llm}}\}\), FedAvg is applied independently to the corresponding \(A\) and \(B\) factors. The resulting \(w_s^{t+1}\) is used to initialize the next training round, and Stages 2–4 are repeated for \(T\) rounds.

\subsection{Computation and Communication Model}
Statistical and system heterogeneity jointly lead to varying client training latency. Local workloads depend on data volumes and modality combinations, while computation capabilities and wireless conditions determine processing and transmission speeds. Assuming sufficient server resources, we neglect aggregation and downlink latency. Thus, the total latency of client \(k\) at round \(t\) is 
\begin{equation}
l_k^t = T_{k,\mathrm{cmp}}^{t} + T_{k,\mathrm{com}}^{t},
\end{equation} 
where \(T_{k,\mathrm{cmp}}^{t}\) and \(T_{k,\mathrm{com}}^{t}\) denote its local computation and uplink communication latency, respectively.

\subsubsection{Computation Model}
The local computational workloads depend on data volumes and modality combinations, which involve different input sizes and encoding and projection paths. The effective data volume of client \(k\) is $$ V_k=D_k\gamma_{\mathcal M_k}, $$ where \(\gamma_{\mathcal M_k}\) captures modality-dependent data volume. Let \(\phi_{\mathcal M_k}\) denote the required computation cycles per unit of effective data,  reflecting modality-dependent processing complexity. The local computation latency is then modeled as
\begin{equation} T_{k,\mathrm{cmp}}^{t} = \frac{V_k\phi_{\mathcal M_k}} {c_k f_k^t},
\end{equation} 
where \(c_k\) denotes the number of available processor cores and \(f_k^t\) denotes the processing frequency of client \(k\) at round \(t\). Hence, \(D_k\), \(\gamma_{\mathcal M_k}\), and \(\phi_{\mathcal M_k}\) capture the heterogeneous local workload, while \(c_kf_k^t\) characterizes the client's available computing capability.

\subsubsection{Communication Model}
Each selected client uploads its updated LoRA parameters with allocated uplink bandwidth \(b_k\) and time-varying signal-to-noise ratio \(\mathrm{SNR}_k^t\) (in dB) at round \(t\). The achievable uplink rate is \cite{zhao2025energy}
\begin{equation} v_k^t = b_k \log_2 \left( 1+10^{\mathrm{SNR}_k^t/10} \right).
\end{equation} 
Let \(S_{\mathrm{LoRA}}\) denote the size (in bits) of the uploaded LoRA parameters, identical across clients under the same LoRA configuration. The uplink latency is
\begin{equation} T_{k,\mathrm{com}}^{t} = \frac{S_{\mathrm{LoRA}}}{v_k^t} .\end{equation} 
Under synchronous FL, round latency is determined by the slowest selected client:
\begin{equation} L^t = \max_{k\in\mathcal A_t} l_k^t.\end{equation}

%Accordingly, client participation exhibits heterogeneous characteristics in both learning and system dimensions: clients differ in their potential training utility due to heterogeneous local data and modality characteristics, while their participation costs differ due to heterogeneous computational workloads, device capabilities, and time-varying wireless conditions. Therefore, client participation must account for both training effectiveness and system efficiency, motivating the dynamic client-selection problem formulated in the following section.
Overall, clients differ in both training utility, due to data and modality heterogeneity, and participation latency, due to computational and wireless heterogeneity. Jointly considering these factors motivates the dynamic client-selection problem formulated next.

\section{Problem Formulation}

Given the heterogeneous multimodal data and dynamic system conditions described above, we aim to determine the participating client subsets over \(T\) FL rounds to improve the final global model performance while satisfying practical latency and wireless resource constraints. The client-selection problem is formulated as
\begin{subequations}
\label{eq:optimization_problem}
\begin{align}
\min_{\{a_k^t\}_{k,t}} \quad
& F(w_s^{T+1}),
\label{eq:objective}
\\
\text{s.t.} \quad
& \sum_{k=1}^{K} a_k^t = n,
\quad \forall t \in \{1,\dots,T\},
\label{eq:selection_cons}
\\
& a_k^t \in \{0,1\},
\quad \forall k,t,
\label{eq:binary_cons}
\\
& L^t \leq \bar{\tau},
\quad \forall t \in \{1,\dots,T\},
\label{eq:latency_cons}
\\
& \sum_{k=1}^{K} a_k^t b_k
\leq B_{\mathrm{total}},
\quad \forall t \in \{1,\dots,T\}.
\label{eq:bandwidth_cons}
\end{align}
\end{subequations} 
The objective in (9a) aims to minimize the final global empirical loss after \(T\) FL rounds. Constraint (9b) requires exactly \(n\) clients to participate in each round, while (9c) defines the binary client-selection variables. \changes{Constraint (9d) requires each communication round to satisfy a prescribed maximum tolerable latency \(\bar{\tau}\), representing an application- or system-level latency requirement and preventing excessive straggler-induced delay.} %Constraint (9d) imposes a maximum tolerable round latency \(\bar{\tau}\) to meet the service requirements of latency-sensitive applications and mitigate excessive straggler-induced delay, 
Constraint (9e) ensures that the aggregate bandwidth requirement of the selected clients does not exceed the total available wireless bandwidth \(B_{\mathrm{total}}\).

Problem (9) is challenging to solve through \changes{offline} optimization, since client training utility depends on heterogeneous data and modality characteristics as well as the evolving global model state, while client participation costs and system feasibility vary with dynamic computation and communication conditions. Moreover, the impact of current selection decisions propagates to subsequent rounds through the evolving global model. Consequently, an effective selection policy should adapt to the evolving training process, account for the long-term impact of current decisions, and satisfy the per-round resource constraints. This motivates the constrained sequential decision-making formulation and learning-based online solution developed in the following section.

\section{CARE-FL Solution}
%To address the optimization problem formulated in Section IV, we propose Constraint-Aware Reinforcement Learning for Efficient Federated MLLM Fine-Tuning (CARE-FL). The framework consists of a deep reinforcement learning (DRL)-assisted dynamic client-selection scheme and an adaptive aggregation mechanism. The former \changes{learns the client-level long-term selection value} and selects clients subject to the latency and bandwidth constraints, while the latter adaptively controls the global update to improve training stability. The two mechanisms are detailed in Sections V-A and V-B, respectively, and their integration into the complete federated MLLM fine-tuning framework is presented in Section V-C.

\changes{To address the optimization problem formulated in Section IV, we propose Constraint-Aware Reinforcement Learning for Efficient Federated MLLM Fine-Tuning (CARE-FL). The framework consists of a Deep Reinforcement Learning (DRL)-assisted dynamic client-selection scheme and an adaptive aggregation mechanism. Rather than directly applying conventional DRL, the former incorporates client-level Q-value decomposition and constraint-aware online learning to enable scalable long-term utility estimation and feasible client selection under latency and bandwidth constraints. The latter adaptively controls potentially unstable global updates to improve training stability. The two mechanisms are detailed in Sections V-A and V-B, respectively, and their integration into the complete federated MLLM fine-tuning framework is presented in Section V-C.}

\subsection{DRL-Assisted Dynamic Client Selection}
%why DRL → MDP + decomposition → state/action/reward/policy.
\changes{Client selection in federated MLLM fine-tuning is inherently sequential, as current participation decisions affect the global model and subsequent client training utilities. To capture such evolving long-term utility, we formulate the client-selection process as an MDP and develop a Constraint-Aware Q-Value Decomposed DDQL Solution.}
%To address the sequential and dynamic client-selection problem formulated in Section IV, we develop a DRL-assisted scheme that learns client participation utilities online from the evolving federated fine-tuning process while accounting for the long-term impact of current decisions. The learned utilities are used together with the latency and bandwidth constraints in Problem (9) to enable feasible online participation decisions. Specifically, we first formulate the client-selection process as a Markov decision process (MDP) with Q-value decomposition and then develop a constraint-aware Double Deep Q-Learning (DDQL) scheme for scalable online client selection.

\subsubsection{MDP Modeling with Q-Value Decomposition} 
We formulate the dynamic client-selection process as an MDP, where the central server acts as the agent and interacts with the FL environment over successive communication rounds. However, directly representing each subset of \(n\) clients as an action results in \(\binom{K}{n}\) possible actions, making conventional joint action-value estimation impractical as \(K\) increases. We therefore incorporate Q-value decomposition into the MDP formulation, approximating the utility of a selected subset through client-level utility estimates. The resulting MDP is defined by the following state, action, reward, and policy objective.

%\textbf{State}: 
\textbf{State:}
    The system state \(s_t\in\mathcal S\) represents the fine-tuning environment at the beginning of round \(t\). We construct the global 
    state from the individual client states as
    \begin{equation}
        s_t = (s_1^t, s_2^t, \dots, s_K^t).
    \end{equation}
    %Each client-specific state \(s_k^t\) is characterized by its model-related information, local data amount, and modality profile. Since the client and global LoRA parameters are high-dimensional, we apply Principal Component Analysis (PCA) to obtain a compact model representation
    \changes{Each client-specific state \(s_k^t\) incorporates information on local and global LoRA parameters, local data amount, and modality profile. To reduce the high dimensionality of LoRA parameters, we employ Gram-matrix-based Principal Component Analysis (PCA) to project their flattened parameter vectors into a shared lower-dimensional space, capturing the dominant parameter variations while reducing the dimensionality of the DRL state representation.}
    \[
        z_k^t = \operatorname{PCA}(w_s^t,w_{k,s}^t),
    \]
    where \(w_s^t\) denotes the current global LoRA 
    parameters and \(w_{k,s}^t\) denotes the latest observed LoRA parameters 
    of client \(k\). The client-specific state is represented as
    \begin{equation}
        s_k^t = (z_k^t, D_k, \mathcal M_k).
    \end{equation}
    %This dimensionality reduction keeps the model-related state information tractable for the DRL agent while incorporating both client-specific and current global model information.\newline \
    \changes{This dimensionality reduction keeps the model-related state information tractable for the DRL agent, while \(D_k\) and \(\mathcal M_k\) explicitly characterize the client's data amount and modality profile.}
    
    \smallskip
    For a client that participated in the previous round, the server refreshes its state using the latest uploaded local parameters. For an unselected client, the server retains the most recently observed state. Therefore, client-state maintenance does not require additional communication from unselected clients.

%\begin{itemize} 
%\item \textbf{Action Space}:
\textbf{Action Space}:
    Under the Q-value decomposition, the utility of each client is evaluated individually. Accordingly, the action \(a_t\in\mathcal A\) is represented by a binary selection vector:
    \begin{equation}
    a_t \in \{0, 1\}^K,
    \end{equation}
    where \(a_k^t=1\) indicates that client \(k\) is selected. \changes{ These client-level selection actions jointly determine the participating subset \(\mathcal U_t=\{k\mid a_k^t=1\}\), with \(|\mathcal U_t|=n\).}

%\begin{itemize}
%    \item \textbf{Reward}: 
\textbf{Reward}: 
    Consistent with the client-level value decomposition, the reward at round \(t\) is represented by the individual utilities observed for the selected clients:
    \begin{equation} r_t = {(r_k^t)}_{k \in \mathcal U_t} = (r_{k_1}^t, r_{k_2}^t, \dots, r_{k_n}^t). \end{equation}

   \changes{To quantify the contribution of each selected client at round \(t\), we first measure the normalized coordinate-wise relative  \(L_1\) divergence between its local LoRA parameters and the global LoRA parameters resulting from round \(t\) and available at the beginning of round \(t+1\), as}
    \begin{equation}
    d(w_{k,s}^{t+1},w_s^{t+1})
    =\frac{1}{d_s}\sum_{i=1}^{d_s}
    \left|
    \frac{w_{k,s}^{t+1}(i)-w_s^{t+1}(i)}
    {\bar w_s^{t+1}(i)}
    \right|,
    \end{equation}
    where \(d_s\) is the number of flattened LoRA parameters and \[\bar w_s^{t+1}(i)=\frac{w_s^{t+1}(i)+\sum_{j\in\mathcal U_t}w_{j,s}^{t+1}(i)}{|\mathcal U_t|+1},\] is the element-wise mean of the accepted global model and the locally updated models of clients selected in round \(t\), used for normalization.
    
    \smallskip
    Based on this divergence and the observed global model performance change, the instantaneous training utility \(\zeta_k^t\), which estimates the contribution of selecting client \(k\) to collaborative model training at round \(t\), is defined as 
    \begin{equation}
    \zeta_k^t = \begin{cases} e^{-d(w_{k,s}^{t+1}, w_s^{t+1})} & \text{if } P_t > P_{t-1}, \\ 1 - e^{-d(w_{k,s}^{t+1}, w_s^{t+1})} & \text{if } P_t \leq P_{t-1}, \end{cases} 
    \end{equation}
    \changes{ where \(P_t\) denotes the server-side validation performance of the accepted global model \(w_s^{t+1}\) resulting from round \(t\). The comparison between \(P_t\) and \(P_{t-1}\) therefore reflects the observed global performance change between two consecutive communication rounds. This performance change determines how the local-global parameter divergence is mapped to instantaneous utility, allowing the reward to reflect both the observed global training outcome and the characteristics of individual client updates. } 

    \smallskip
    
    \changes{The exponential mapping transforms the nonnegative divergence into a smooth and bounded utility. The two complementary branches provide opposite monotonic preferences over client divergence while maintaining the same utility range. When the global performance improves, \(e^{-d}\) assigns higher utility to clients with smaller local-global divergence, favoring updates that are more consistent with the improving global model. Otherwise, \(1-e^{-d}\) assigns higher utility to clients with larger divergence, encouraging exploration of clients with more distinct update characteristics. Moreover, the exponential mapping gradually saturates as the divergence increases, limiting the influence of excessively large parameter deviations on the reward.}
 
    The final reward includes historical feedback through
    \begin{equation}
    r_k^t = \lambda \zeta_k^t + (1-\lambda) \zeta_k^{t-1}. 
    \end{equation}
    The historical utility is initialized as \(\zeta_k^0=0\) for all clients and remains unchanged in rounds when the client is not selected. The parameter \(\lambda\in[0,1]\) controls the contribution of the current utility. This temporal smoothing reduces short-term reward fluctuations while retaining information about each client's recent contribution.
    
%\end{itemize}

%\begin{itemize}
    %\item \textbf{Policy Optimization}: 
\textbf{Policy Optimization}: 
    The DRL agent seeks a policy \(\pi^*\) that maximizes the expected cumulative discounted reward over \(T\) fine-tuning rounds:
    \begin{equation}
    \pi^* = \arg \max_{\pi} \mathbb{E} \left[ \sum_{t=1}^{T} \gamma^{t-1}R(s_t, a_t) \right], 
    \end{equation}
    where $\gamma \in (0, 1]$ is the reward discount factor. The aggregate reward is defined as the sum of the individual rewards of the selected clients:
    \begin{equation} R(s_t,{a}_t) = \sum_{k=1}^{K} a_k^t r_k^t = \sum_{k \in \mathcal U_t} {r_k^t}. \end{equation}
   Therefore, the policy objective promotes client subsets with high cumulative
   utility over the training horizon.
%\end{itemize}

\subsubsection{Constraint-Aware Q-Value Decomposed DDQL Solution}
To solve the above MDP, we develop a constraint-aware Q-value decomposed DDQL scheme that integrates three components: a decomposed Q-value network for scalable client-level long-term utility estimation, a modified DDQL update for online learning of these utilities from fine-tuning feedback, and constraint-aware \(\epsilon\)-greedy execution for selecting feasible client subsets under the latency and bandwidth constraints.   
\begin{figure}[htbp]
    \centering
    \includegraphics[width=1\linewidth]{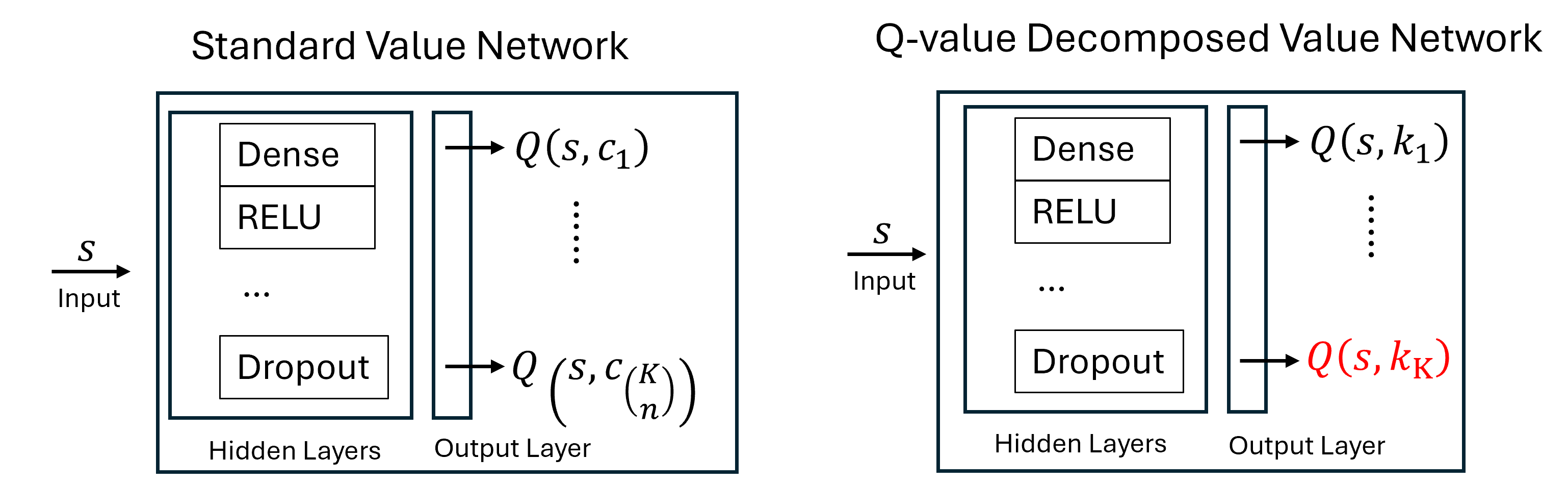}
    \caption{Comparison of DRL Q-value network architectures. The conventional joint-action value network (left) outputs \(\binom{K}{n}\) Q-values, whereas the proposed Q-value decomposed network (right) outputs \(K\) client-level utility estimates.}
    \label{fig:network_comparison}
\end{figure}

%\begin{itemize}   
    %\item \textbf{Decomposed Q-Value Network}: 
\textbf{Decomposed Q-Value Network}: 
    As illustrated in Fig. 2, directly estimating \(Q_\theta(s_t,a_t)\) for every possible client subset requires an output dimension of \(\binom{K}{n}\), \changes{which becomes impractical as \(K\) increases. Since the aggregate reward in (18) is constructed from the individual rewards of the selected clients, it naturally provides client-level feedback on the contribution of each selected client. Motivated by this client-level reward structure, we approximate the long-term value of a selected subset through an additive combination of client-level Q-values. This design follows the general principle of value-function factorization, where a joint action value is represented through individual utility functions \cite{sunehag2017value, rashid2020monotonic, huang2025qvf}. Accordingly, instead of estimating Q-values for all client subsets, the proposed network outputs \(K\) client-level Q-value estimates and approximates the value of a selected subset by}
    %To avoid this combinatorial representation, our network instead outputs \(K\) client-level utility estimates and approximates the value of a selected subset by
    \begin{equation} Q_\theta(s_t, a_t) \approx \sum_{k=1}^K a_k^t Q_\theta(s_t, k) =\sum_{k \in \mathcal U_t} Q_\theta(s_t, k),  \end{equation}
    where \(Q_\theta(s_t,k)\) denotes the estimated long-term utility associated with selecting client \(k\) under state \(s_t\), learned from its participation in jointly selected client subsets. Consequently, the Q-network output dimension is reduced from \(\binom{K}{n}\) to \(K\). Based on these client-level utilities, client selection can be expressed as 
    \begin{equation} a_t^* = \arg \max_{a \in \mathcal{A}} \sum_{k=1}^K a_k^t \cdot Q_\theta(s_t, k),\quad \text{s.t. } \sum_{k=1}^K a_k^t = n. \end{equation}
    
%\end{itemize}

%\begin{itemize}
    %\item \textbf{Modified DDQL Optimization}: 
\textbf{Modified DDQL Optimization}: 
    To train the decomposed Q-value network, we adopt the online-target network architecture of DDQL and modify its target construction to accommodate client-level long-term utility estimation. Specifically, the online network \(Q_\theta\) estimates the client long-term utilities and is optimized during training, while the target network \(Q_{\theta^-}\) is used to construct stable target values.
    
    \smallskip
\changes{
    \noindent Individual Target Value Construction: Conventional DDQL \cite{van2016deep} uses the online network for action selection and the target network for evaluation, yielding
    \begin{equation}
    Y_t = r_t + \gamma Q_{\theta^-}
    \left(s_{t+1},
    \arg\max_{a'} Q_\theta(s_{t+1},a')\right).
    \end{equation}
    In our decomposed framework, instead of constructing a target for the joint client-subset action, we construct an individual temporal-difference target for each participating client by bootstrapping its future utility estimate at the next state:
    %rather than performing next-action maximization over the combinatorial joint action space as in conventional DDQL, we directly bootstrap the future utility of client \(k\) under the next state:
    \begin{equation}Y_k^t = r_k^t + \gamma Q_{\theta^-}(s_{t+1}, k), \end{equation}
    where \(r_k^t\) is the immediate client-level reward and \(Q_{\theta^-}(s_{t+1},k)\) provides the bootstrapped estimate of the future training utility of client \(k\). This modified target is consistent with the client-level decomposition in (19), enabling each participating client's utility to be updated individually from the observed fine-tuning feedback without constructing a target over the combinatorial client-subset action space.
}

    Decomposed Loss Function: Based on the individual targets, the online network is optimized by minimizing the client-level Q-value estimation errors:
    \begin{equation}
    L(\theta)
    =
    \mathbb{E}\left[
    \sum_{k\in\mathcal U_t}
    \left(Y_k^t-Q_\theta(s_t,k)\right)^2
    \right].
    \end{equation}
    The additive loss enables direct credit assignment to the participating clients, while the target network is periodically synchronized with the online network every \(I\) rounds, where \(I\) denotes the target-network update interval. The resulting learning procedure is summarized in \textbf{Algorithm 1}.
   
%\end{itemize}

%\begin{itemize}
    %\item \textbf{Constraint-Aware $\epsilon$-greedy Online RL Client Selection}: 
\textbf{Constraint-Aware $\epsilon$-greedy Online RL Client Selection}:
    The learned client-level Q-values provide utility estimates for client selection. However, directly selecting the clients with the highest Q-values may violate the bandwidth and latency constraints. We therefore develop a constraint-aware \(\epsilon\)-greedy strategy that converts the learned utilities into feasible client subsets while maintaining exploration during online learning. 
    
    Specifically, the exploitation step solves
    \begin{subequations} \label{eq:client_selection_optimization_problem}
    \begin{align}
    a_t^* = \arg \max_{a \in \mathcal{A}} & \sum_{k=1}^K a_k^t \cdot Q_\theta(s_t, k) ,\label{eq:obj_func} \\
    \text{s.t.} \quad & \sum_{k=1}^K a_k^t \cdot b_k \leq B_{\mathrm{total}} ,\label{eq:const_bw} \\
    & \max_{k\in \mathcal U_t} l^t_k \leq \bar{\tau} ,\label{eq:const_lat} \\
    & \sum_{k=1}^K a_k^t = n .\label{eq:const_num}
    \end{align}
    \end{subequations}
    This constrained Top-\(n\) problem selects clients with high estimated long-term utilities while ensuring system feasibility and can be solved as an integer linear program (ILP) for the considered client scale. Equation (24) executes the learned Q-values under constraints rather than globally solving the original sequential optimization problem. %Note that (24) represents a constraint-aware execution of the learned Q-values rather than a globally optimal solution to the original sequential optimization problem. 

    To retain exploration, we integrate the constrained Top-\(n\) selection into an \(\epsilon\)-greedy mechanism. At round \(t\), \(\lfloor(1-\epsilon)n\rfloor\) clients are first selected according to their learned Q-values subject to (24). The remaining slots are filled by random exploration, where a candidate client is admitted only if the resulting subset remains feasible under the residual bandwidth and latency constraints. The exploration rate \(\epsilon\) gradually decays during training, shifting the policy from exploration toward exploitation. The complete procedure is summarized in \textbf{Algorithm 2}.

%\end{itemize}

\begin{algorithm}[t]
    \caption{Client-Level DDQL Learning}
    \label{alg:DDQL}
    \begin{algorithmic}[1]
    \Procedure{UpdateAgent}{$\mathcal{B}, \theta, \theta^-, \gamma, t, I$}
        \State Sample a minibatch of $N$ transitions $\{(s_{t,j}, k_j, r_{k,j}, s_{t+1,j})\}_{j=1}^N$ from $\mathcal{B}$
        \For{each transition $j \in \{1, \dots, N\}$}
            \State $Y_j \gets r_{k,j} + \gamma Q_{\theta^-}(s_{t+1,j}, k_j)$
        \EndFor
        \State $L(\theta) \gets \frac{1}{N} \sum_{j=1}^N (Y_j - Q_\theta(s_{t,j}, k_j))^2$
        \State Update $\theta$ by performing gradient descent on $L(\theta)$
        \If{$t \pmod{I} == 0$}
            \State $\theta^- \gets \theta$
        \EndIf
    \EndProcedure
    \end{algorithmic}
\end{algorithm}

\begin{algorithm}[t]
\caption{Constraint-Aware $\epsilon$-Greedy Client Selection}
\label{alg:constrained_egreedy_selection}
\begin{algorithmic}[1]

\Procedure{ConstraintSelect}{$Q_{\theta},s_t,\epsilon,n,B_{\mathrm{total}},\bar{\tau}$}

    \Statex \hspace{1em}\textbf{// Stage 1: Q-guided exploitation}
    \State $n_{\mathrm{q}}\leftarrow\lfloor(1-\epsilon)n\rfloor$
    \State $\mathcal{U}_{\mathrm{q}}\leftarrow
    \textsc{ConstraintTop}(Q_{\theta},s_t,n_{\mathrm{q}},B_{\mathrm{total}},\bar{\tau})$

    \Statex \hspace{1em}\textbf{// Stage 2: Random exploration}
    \State $n_{\mathrm{r}}\leftarrow n-n_{\mathrm{q}}$
    \State $B_{\mathrm{rem}}\leftarrow
    B_{\mathrm{total}}-\sum_{k\in\mathcal{U}_{\mathrm{q}}}b_k$
    \State $\mathcal{U}_{\mathrm{r}}\leftarrow
    \textsc{ConstraintRandom}(n_{\mathrm{r}},B_{\mathrm{rem}},\bar{\tau},\mathcal{U}_{\mathrm{q}})$
    \State $\mathcal{U}_t\leftarrow
    \mathcal{U}_{\mathrm{q}}\cup\mathcal{U}_{\mathrm{r}}$
    \State \Return $\mathcal{U}_t$

\EndProcedure

\Statex \hrulefill

\Procedure{ConstraintTop}{$Q_{\theta},s_t,n',B,\tau$}

    \State $\mathcal{K}_{\mathrm{feas}}\leftarrow
    \{k\in\mathcal{K}\mid l_k^t\leq\tau\}$

    \State Solve $a_t^\star=\arg\max
    \sum_{k\in\mathcal{K}_{\mathrm{feas}}}
    a_k^tQ_{\theta}(s_t,k)$

    \Statex \hspace{2em}
    $\mathrm{s.t.}\ 
    \sum_{k\in\mathcal{K}_{\mathrm{feas}}}a_k^tb_k\leq B,\quad
    \sum_{k\in\mathcal{K}_{\mathrm{feas}}}a_k^t=n'$

    \State \Return
    $\{k\in\mathcal{K}_{\mathrm{feas}}\mid a_k^{t\star}=1\}$

\EndProcedure

\Statex \hrulefill

\Procedure{ConstraintRandom}{$n',B,\tau,\mathcal{U}_{\mathrm{excl}}$}

    \State $\mathcal{U}_{\mathrm{tmp}}\leftarrow\emptyset$

    \State $\mathcal{C}\leftarrow
    \{k\in\mathcal{K}\setminus\mathcal{U}_{\mathrm{excl}}
    \mid l_k^t\leq\tau\}$

    \While{$|\mathcal{U}_{\mathrm{tmp}}|<n'$
    \textbf{ and } $\mathcal{C}\neq\emptyset$}

        \State Randomly sample $k\in\mathcal{C}$

        \If{$b_k\leq B$}
            \State $\mathcal{U}_{\mathrm{tmp}}\leftarrow
            \mathcal{U}_{\mathrm{tmp}}\cup\{k\}$
            \State $B\leftarrow B-b_k$
        \EndIf

        \State $\mathcal{C}\leftarrow\mathcal{C}\setminus\{k\}$

    \EndWhile

    \State \Return $\mathcal{U}_{\mathrm{tmp}}$

\EndProcedure

\end{algorithmic}
\end{algorithm}

\subsection{Adaptive Aggregation}
Although the proposed RL scheme enables adaptive client selection, exploratory actions and heterogeneous client updates may occasionally produce unstable global updates, particularly during the early policy-learning stage. To improve training stability, we employ SlowFed, an adaptive aggregation mechanism that controls the magnitude of the global update according to the observed model performance.

At round \(t\), the server first obtains a trial model using the standard FedAvg update
\begin{equation}w_{s,\mathrm{avg}}^{t+1}=w_s^t+\Delta w_s^t, \end{equation} 
\changes{where \(\Delta w_s^t\) denotes the aggregated LoRA update obtained from the selected clients. The server then evaluates the trial model on a server-side validation set. If its performance degradation relative to the current global model does not exceed a predefined threshold \(\Gamma\), i.e., $$ \operatorname{Acc}(w_{s,\mathrm{avg}}^{t+1}) \geq \operatorname{Acc}(w_s^t)-\Gamma, $$ the trial update is accepted and \(w_s^{t+1}=w_{s,\mathrm{avg}}^{t+1}\). Otherwise, SlowFed triggers a conservative recovery procedure by scaling the trial update with a server-side learning rate \(\delta_i\).}
\begin{equation}
w_s^{t+1,(i)}
=
(1-\delta_i)w_s^t
+\delta_i w_{s,\mathrm{avg}}^{t+1}
=
w_s^t+\delta_i\Delta w_s^t.
\end{equation}
Progressively smaller \(\delta_i\) values are examined until the stability condition is satisfied. If none of the recovery attempts satisfies the condition, the server retains the previous global model, i.e., \(w_s^{t+1}=w_s^t\). \changes{The complete SlowFed aggregation procedure is summarized in \textbf{Stage 4 of Algorithm 3}}.

\begin{comment}
\begin{algorithm}[t]
\caption{Adaptive Aggregation (SlowFed)}
\label{alg:SlowFed}
\begin{algorithmic}[1]
\Procedure{SlowFedAggregation}
{$w_s^t, \Delta w_s^t, \mathrm{Acc}_t, \Gamma, R, \{\delta_1, \dots, \delta_R\}$}

    \Statex \hspace{1em} \textbf{// Step 1: Initial Trial Update}
    \State $w_{s,\mathrm{avg}}^{t+1}
    \gets w_s^t + \Delta w_s^t$

    \State $\mathrm{Acc}_{t+1}
    \gets \text{Eval}(w_{s,\mathrm{avg}}^{t+1})$

    \Statex \hspace{1em} \textbf{// Step 2: Stability Check and Recovery Loop}
    \If{$\mathrm{Acc}_{t+1}
    < \mathrm{Acc}_t - \Gamma$}

        \State $i \gets 1$

        \While{$(\mathrm{Acc}_{t+1}
        < \mathrm{Acc}_t - \Gamma)$
        \textbf{ and } $i \le R$}

            \State $w_s^{t+1}
            \gets
            (1-\delta_i)w_s^t
            +\delta_i w_{s,\mathrm{avg}}^{t+1}$

            \State $\mathrm{Acc}_{t+1}
            \gets \text{Eval}(w_s^{t+1})$

            \State $i \gets i+1$

        \EndWhile

    \Else

        \State $w_s^{t+1}
        \gets w_{s,\mathrm{avg}}^{t+1}$

    \EndIf

    \Statex \hspace{1em} \textbf{// Step 3: Fallback Protection}
    \If{$\mathrm{Acc}_{t+1}
    < \mathrm{Acc}_t - \Gamma$}

        \State $w_s^{t+1} \gets w_s^t$

    \EndIf

    \State \Return $w_s^{t+1}$

\EndProcedure

\end{algorithmic}
\end{algorithm}
\end{comment}

\begin{algorithm}[t]
\caption{CARE-FL Framework}
\label{alg:carefl}
\begin{algorithmic}[1]

\Require Clients $\mathcal{K}$, rounds $T$, selected clients $n$,
$B_{\mathrm{total}}$, $\bar{\tau}$, $\Gamma$,
$\{\delta_1,\ldots,\delta_R\}$.
\Ensure Global LoRA parameters $w_{\mathrm{s}}^{T+1}$.

\Statex \textbf{// Stage 1: Initialization}
\State Initialize $w_{\mathrm{s}}^1$, $Q_\theta$, $Q_{\theta^-}$,
and replay buffer $\mathcal{B}$.
\State All clients perform one-time warm-up to initialize $s_1$.

\For{$t=1$ to $T$}

    \Statex \textbf{// Stage 2: Constraint-Aware Client Selection}
    \State $\mathcal{U}_t \gets
    \Call{ConstraintSelect}
    {Q_\theta,s_t,\epsilon,n,B_{\mathrm{total}},\bar{\tau}}$

    \Statex \textbf{// Stage 3: Local Fine-Tuning}
    \State Each selected client $k \in \mathcal{U}_t$ locally fine-tunes $w_{\mathrm{s}}^t$ and uploads $\Delta w_{k,\mathrm{s}}^t$.

    \Statex \textbf{// Stage 4: SlowFed Aggregation}
    \State $\displaystyle
    \Delta w_{\mathrm{s}}^t \gets
    \sum_{k\in\mathcal{U}_t}
    \frac{|D_k|}{\sum_{j\in\mathcal{U}_t}|D_j|}
    \Delta w_{k,\mathrm{s}}^t$
    \State $w_{\mathrm{s,avg}}^{t+1}
    \gets w_{\mathrm{s}}^t+\Delta w_{\mathrm{s}}^t$
    \State $\mathrm{Acc}_{t+1}
    \gets \Call{Eval}{w_{\mathrm{s,avg}}^{t+1}}$

    \If{$\mathrm{Acc}_{t+1}<\mathrm{Acc}_t-\Gamma$}
        \For{$i=1$ to $R$}
            \State $w_{\mathrm{s}}^{t+1}
            \gets w_{\mathrm{s}}^t+\delta_i\Delta w_{\mathrm{s}}^t$
            \State $\mathrm{Acc}_{t+1}
            \gets \Call{Eval}{w_{\mathrm{s}}^{t+1}}$
            \If{$\mathrm{Acc}_{t+1}\geq\mathrm{Acc}_t-\Gamma$}
                \State \textbf{break}
            \EndIf
        \EndFor
        \If{$\mathrm{Acc}_{t+1}<\mathrm{Acc}_t-\Gamma$}
            \State $w_{\mathrm{s}}^{t+1}\gets w_{\mathrm{s}}^t$
        \EndIf
    \Else
        \State $w_{\mathrm{s}}^{t+1}\gets w_{\mathrm{s,avg}}^{t+1}$
    \EndIf

    \Statex \textbf{// Stage 5: State and RL Update}
    \State Construct $s_{t+1}$ and compute
    $\{r_t^k\}_{k\in\mathcal{U}_t}$.
    \State Store $\{(s_t,k,r_t^k,s_{t+1})\}_{k\in\mathcal{U}_t}$
    in $\mathcal{B}$.
    \State \Call{UpdateAgent}
    {$\mathcal{B},\theta,\theta^-,\gamma,t,I$}

\EndFor

\State \Return $w_{\mathrm{s}}^{T+1}$

\end{algorithmic}
\end{algorithm}

\subsection{Federated MLLM LoRA Fine-Tuning Framework}

With the proposed RL-assisted client selection and adaptive aggregation 
mechanisms, we now integrate the above components into the complete 
federated MLLM LoRA fine-tuning framework. The overall training procedure, summarized in \textbf{Algorithm 3}, consists of the following five stages.

\textbf{Stage 1: Initialization.}
The server initializes the global LoRA parameters,
decomposed Q-networks, and replay buffer $\mathcal{B}$. The initial validation accuracy is set to $\mathrm{Acc}_1=0$. Each client performs a one-time local warm-up update
to construct the initial state $s_1$, reflecting local
data and modality heterogeneity. This warm-up is
excluded from the subsequent constrained training rounds.

For each round $t=1,\ldots,T$, CARE-FL performs:

\textbf{Stage 2: Constraint-Aware Client Selection.}
Given $s_t$, the server selects $\mathcal{U}_t$ using
the constraint-aware $\epsilon$-greedy strategy in
\textbf{Algorithm~2} and distributes the global LoRA parameters to the selected clients.

\textbf{Stage 3: Local Fine-Tuning and Upload.}
Selected clients fine-tune their parameters
locally and upload the updates.

\textbf{Stage 4: Adaptive Aggregation.} The server aggregates local updates weighted by data
size. SlowFed then adaptively controls this update to produce the next global LoRA parameters \(w_s^{t+1}\), mitigating potentially unstable global updates.

\textbf{Stage 5: State and RL Update.} The server computes the client-level rewards \(r_k^t\), constructs the next state \(s_{t+1}\), and stores the transitions \((s_t,k,r_k^t,s_{t+1})\) for \(k\in\mathcal U_t\) in \(\mathcal B\). The decomposed Q-network is then updated online according to the modified DDQL procedure in \textbf{Algorithm 1}, using minibatches sampled from the experience replay buffer \(\mathcal B\). The resulting \(s_{t+1}\) becomes the state for the next communication round.

%Stages 2--5 are repeated throughout the \(T\) communication rounds, jointly coupling constraint-aware client selection, federated LoRA fine-tuning, adaptive aggregation, and online RL learning.

\section{Experiments}
In this section, we present the empirical validation of the proposed CARE-FL framework. We first outline the experimental setup, including model architecture, dataset specifications, and baseline comparisons. Subsequently, we demonstrate the effectiveness of our method in terms of both global model accuracy and system efficiency under resource-constrained MEC environments. 

\subsection{Experimental Settings}
\textbf{MLLM Model and PEFT Configuration}: We adopt MiniCPM-V 2.0 \cite{yao2024minicpm}, a 2.8B-parameter MLLM comprising a SigLIP vision encoder, a MiniCPM language backbone, and a Perceiver Resampler. %The resampler compresses visual representations into 64 tokens before they are processed by the language model.

The vision encoder and resampler are frozen, and LoRA is applied to the self-attention modules (\(q_{\mathrm{proj}}, k_{\mathrm{proj}}, v_{\mathrm{proj}}, o_{\mathrm{proj}}\)) of the LLM backbone. By default, the LoRA rank and scaling factor are set to \(r=8\) and \(\alpha=8\), respectively, resulting in approximately 5.9 million ($\sim 0.17\%$) trainable parameters.

\textbf{Dataset, Task, and Heterogeneous Data Partitioning}: We use the CrisisMMD dataset\cite{alam2018crisismmd}, a multimodal disaster-response dataset containing more than 16,000 annotated image-text pairs collected from seven major natural disasters. Following the preprocessing methodology of \cite{ofli2020analysis}, %we retain samples for which the image and text annotations are consistent. To reduce class imbalance, semantically overlapping categories are merged: injured or dead and missing or found are combined into Affected Individuals, while vehicle damage is incorporated into Infrastructure and Utility Damage. 
\changes{the resulting dataset contains five humanitarian categories: Affected Individuals, Infrastructure Damage, Rescue Efforts, Other Relevant Information, and Not Humanitarian, with 6,126 training samples, 998 validation samples, and 998 test samples.}

We formulate the task as multimodal instruction tuning, where the MLLM predicts a humanitarian category from an image $X_v$ and textual input $X_q$ containing the sample text and classification instruction.
The language modeling loss is
%We formulate the downstream task as multimodal instruction tuning. Given an image \(X_v\) and its associated textual input \(X_q\), which contains both the sample-specific text and an instruction to select from the five candidate humanitarian categories, the MLLM generates a textual response \(Y\) corresponding to the ground-truth humanitarian category. The training objective follows the causal language modeling loss:
\begin{equation} \mathcal{L} = -\sum_{i=1}^{N_y} \log P(y_i\mid X_v,X_q,y_{<i};\theta), \end{equation}
where \(N_y\) is the target response length, \(y_i\) denotes the \(i\)-th token of \(Y\), \(y_{<i}\) denotes the preceding target tokens, and \(\theta\) denotes the trainable LoRA parameters.

%The processed dataset is stratified into training (70\%), validation (15\%), and testing (15\%) sets. Only the training set is distributed among the clients for federated fine-tuning, while the validation and test sets are retained at the server for validation and final performance evaluation, respectively. To simulate statistical heterogeneity, the training data are partitioned among \(K=50\) clients using a Dirichlet distribution with concentration parameter \(\alpha=0.1\), resulting in heterogeneous label distributions and local data quantities. Modality heterogeneity is further introduced by assigning clients to three modality configurations: text-only, image-only, and full image-text modalities. Unless otherwise specified, the three configurations are approximately equally represented across the client population. Consequently, clients differ in both their local data distributions and modality availability.

The training set is partitioned among (K=50) clients using a Dirichlet distribution with \(\alpha=0.1\) to introduce heterogeneous label distributions and data quantities, while validation and test sets remain at the server. Modality heterogeneity is introduced through approximately equal proportions of text-only, image-only, and image-text clients.

\textbf{Baselines}: We compare the proposed framework with the following client-selection methods: 

\textit{\underline {FedAvg}}: Randomly selects \(n\) clients per round.

\changes{\textit{\underline {FAVOR}} \cite{wang2020optimizing}: An accuracy-oriented RL method using a Deep Q-Network (DQN) with model weights as state inputs.} %Notably, its selection policy is trained on a Top-1 basis and greedily extended to select $n$ clients, ignoring system latency and hardware heterogeneity.}

\changes{\textit{\underline {FLASH-RL}} \cite{bouaziz2023flash}: A system-aware RL method incorporating hardware characteristics into client states and employing client-specific rewards.}

%A system-aware RL baseline that incorporates hardware properties (e.g., CPU frequency) into client-specific states and utilizes client-specific rewards.} %Its goal is to jointly optimize both model accuracy and training latency. However, a key limitation is that it balances these conflicting objectives within a single joint reward function without a stable weighting mechanism, often leading to inconsistent trade-offs between performance and system efficiency compared to our approach.

Under resource constraints, all methods follow their original selection policies while sequentially admitting clients that satisfy the same bandwidth and latency constraints, until \(n\) clients are selected.
%For comparisons under resource constraints, we impose the same per-round bandwidth and latency requirements on all methods. For each baseline, candidate clients are considered according to its original selection policy and sequentially admitted only when their inclusion preserves both constraints, until \(n\) feasible clients are selected. This ensures that all methods operate under the same resource constraints while preserving their respective client-selection criteria.

\textbf{Evaluation Metrics}: To provide a comprehensive assessment of CARE-FL, we utilize the following key metrics: 

\textit{\underline {Test Accuracy}}: We measure the Top-1 accuracy of the global model on the held-out test set, where generated responses are mapped to the corresponding humanitarian categories.

\textit{\underline {Latency Usage}}: We calculate the per-round latency determined by the slowest selected client.

\textit{\underline {Total Bandwidth Usage}}: Total bandwidth usage is calculated as the summation of the individual bandwidth for all selected clients within that round. 

The latter two metrics are used to evaluate system efficiency and verify satisfaction of the latency and bandwidth constraints.

\textbf{Implementation and Training Parameters}: 
CARE-FL is implemented using PyTorch and Hugging Face. Local fine-tuning employs AdamW (learning rate of \(2\times10^{-5}\), weight decay 0.1, \(\beta_2=0.95\)) with cosine learning rate decay and 1\% warm-up. We use a batch size of 1, BF16 precision, gradient checkpointing, a maximum sequence length of 1200 tokens, and up to nine image slices. Each selected client performs one local epoch per round. Table I summarizes the FL, LoRA, RL, and SlowFed hyperparameters. The replay buffer stores client-level transitions, with each DDQL update using all available samples.

%The proposed framework is implemented using PyTorch and the Hugging Face ecosystem. For local fine-tuning, we use the AdamW optimizer with a learning rate of \(2\times10^{-5}\), weight decay of \(0.1\), and \(\beta_2=0.95\). The learning rate follows a cosine decay schedule with a 1\% warm-up ratio. The per-device training batch size is set to 1, and gradient checkpointing and bfloat16 (BF16) precision are employed to reduce memory consumption. The maximum sequence length is set to 1200 tokens, with at most nine image slices for visual inputs.

%Federated fine-tuning is conducted for \(T=50\) communication rounds with \(K=50\) clients, of which \(n=10\) are selected in each round. Each selected client performs \(E=1\) local training epoch before uploading its updated LoRA parameters to the server. The detailed FL, LoRA, RL, and adaptive aggregation hyperparameter settings are summarized in Table I.

\begin{table}[t]
\centering
\caption{Experimental Parameter Settings}
\label{tab:parameter_settings}
\renewcommand{\arraystretch}{1.05}
\setlength{\tabcolsep}{4pt}
\begin{tabular}{l l l}
\hline
\textbf{Category} & \textbf{Parameter} & \textbf{Value} \\
\hline
FL
& $(K,n,T,E)$ & $(50,10,50,1)$ \\

LoRA
& $(r,\alpha)$ & $(8,8)$ \\

Local training
& Learning rate & $2\times10^{-5}$ \\

\hline
RL
& Discount factor $\gamma$ & $0.9$ \\
& Exploration rate $\epsilon$ & $0.9 \rightarrow 0.2$ \\
& Exploration decay & Linear over $T$ rounds \\
& Reward smoothing factor $\lambda$ & $0.6$ \\
& Replay buffer capacity & $1000$ \\
& DDQL update batch & Entire current buffer \\
& DDQL learning rate & $0.01$ \\
& Target network update interval & $5$ rounds \\
& Gradient steps per DDQL update & $30$ \\

\hline
SlowFed
& Performance threshold $\Gamma$ & $3\%$ \\
& Recovery factors $\boldsymbol{\delta}$ 
& $\{0.1,0.05,0.025\}$ \\
& Max. recovery attempts $R$ & $3$ \\
\hline

\end{tabular}
\end{table}

\textbf{Experimental Platform and Heterogeneous Distributed System}: Experiments are conducted on two NVIDIA RTX 4090 (24 GB) GPUs, centrally emulating a wireless FL system with 50 heterogeneous clients. Client profiles differ in bandwidth, channel conditions, computational resources, and modality-dependent workloads. Following [52], client SNR and CPU frequency vary across rounds to simulate dynamic computation and communication conditions.

\subsection{Overall Accuracy Performance}
Unless otherwise specified, all reported accuracy curves are averaged over three independent runs to reduce run-to-run variability. In the unconstrained setting, FLASH-RL is evaluated using both its default configuration, \(\alpha_1=1,\alpha_2=1\), where \(\alpha_1\) and \(\alpha_2\) weight statistical and system utility in its reward, and an additional configuration with \(\alpha_1=1,\alpha_2=0\). The latter excludes system utility, enabling evaluation of selection based solely on statistical utility without explicit resource constraints.

Under the unconstrained setting, Fig. 3 shows that CARE-FL reaches high accuracy earlier and maintains the strongest late-round performance. After approximately 20 communication rounds, its test accuracy generally remains above 80\%, whereas the baselines show lower accuracy and greater fluctuations. Table II further shows that CARE-FL achieves the highest final-round mean accuracy of 82.56\%, with a late-stage accuracy range of only 1.91 percentage points. These results indicate improved convergence and relatively consistent accuracy across runs.

This advantage aligns with CARE-FL's design. \changes{The state representation incorporates client-specific modality information, allowing the learned utilities to capture modality heterogeneity and evolving client contributions.} Client-level Q-value decomposition replaces combinatorial joint-action representation with individual utility estimates, enabling scalable learning while preserving long-term client evaluation. Without explicit resource constraints, the resulting policy prioritizes clients by learned long-term training utility, helping explain the stronger convergence in Fig. 3 and Table II.

%The client-level Q-value decomposition reduces the combinatorial joint-action representation to individual client-level utility estimates, \changes{ enabling more efficient and scalable utility learning while retaining effective long-term client evaluation.}In the absence of explicit resource constraints, the resulting policy can therefore prioritize clients according to their learned long-term training utility. This provides a mechanism-level explanation for the stronger convergence behavior observed in Fig. 3 and  Table II.

\begin{figure}[!t]
    \centering
    \includegraphics[width=\linewidth]
    {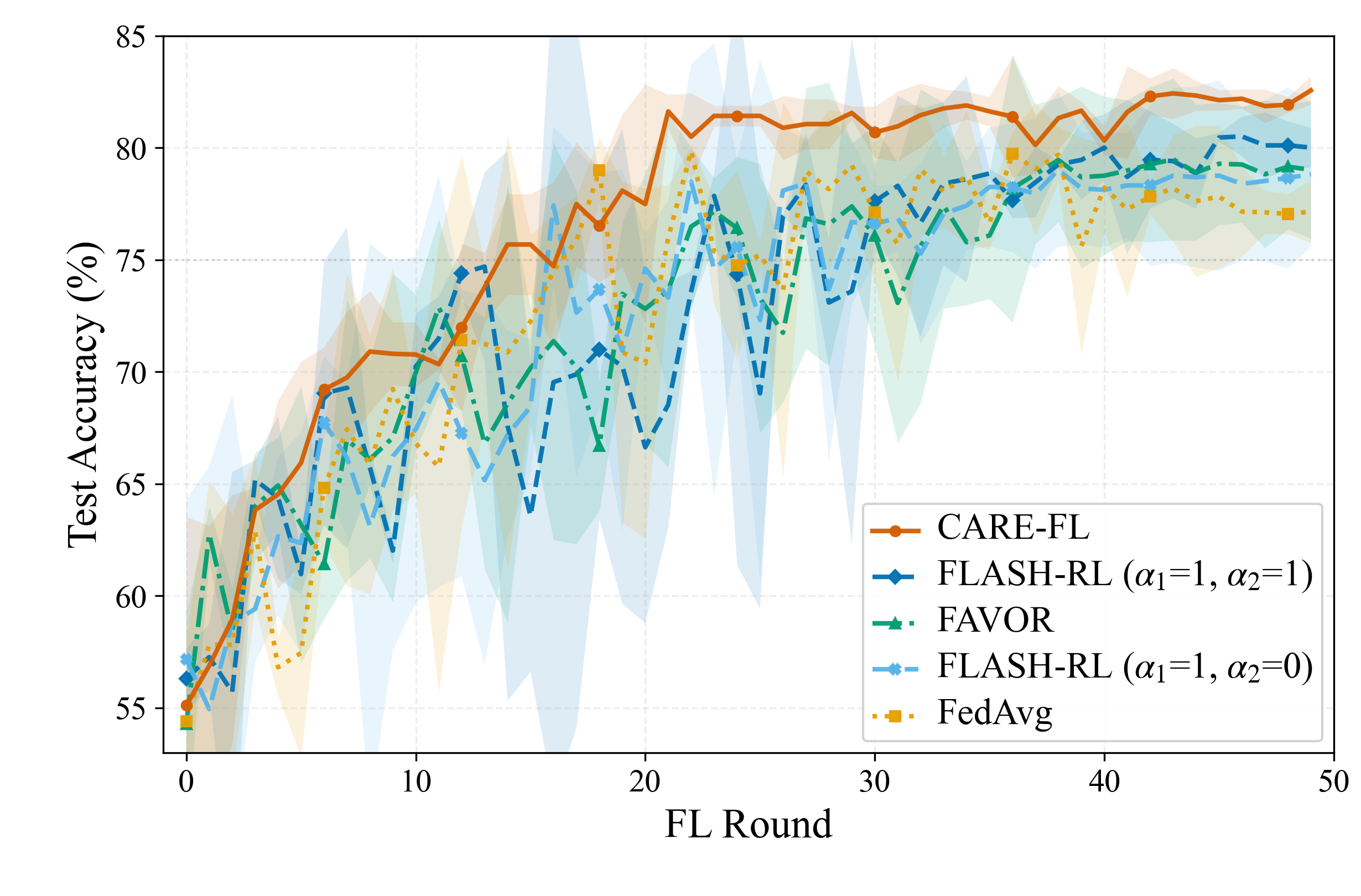}

    \caption{Test accuracy convergence without resource constraints.
    The curves and shaded regions represent the mean test accuracy
    and its variability across independent runs, respectively.}
    \label{fig:accuracy_unconstrained}
\end{figure}

\begin{table}[t]
    \centering
    \caption{Final-round accuracy and late-stage variation without
    resource constraints.}
    \label{tab:final_stage_unconstrained}

    \small
    \setlength{\tabcolsep}{3.5pt}
    \renewcommand{\arraystretch}{1.12}

    \begin{tabular}{lcccc}
        \hline
        \textbf{Method}
        & \textbf{Final Mean}
        & \textbf{Last-3 Min}
        & \textbf{Last-3 Max}
        & \textbf{Range} \\
        & \textbf{(\%)}
        & \textbf{(\%)}
        & \textbf{(\%)}
        & \textbf{(pp)} \\
        \hline
        CARE-FL
        & 82.56 & 81.26 & 83.17 & 1.91 \\
        FLASH-RL $(1,1)$
        & 80.01 & 79.36 & 80.86 & 1.50 \\
        FAVOR
        & 79.02 & 76.05 & 81.86 & 5.81 \\
        FLASH-RL $(1,0)$
        & 78.82 & 76.35 & 81.96 & 5.61 \\
        FedAvg
        & 77.16 & 75.75 & 78.06 & 2.31 \\
        \hline
    \end{tabular}

    \vspace{1mm}
    \begin{minipage}{0.98\linewidth}
        \footnotesize
        Final Mean denotes the mean test accuracy at the final
        communication round. Last-3 Min and Last-3 Max denote the
        minimum and maximum test accuracies observed over the last
        three communication rounds across independent runs. Range is
        their difference in percentage points (pp).
    \end{minipage}
\end{table}

\begin{figure*}[!t]
    \centering

    % ---------------------------------------------------------
    % (a) Test accuracy convergence
    % ---------------------------------------------------------
    \begin{subfigure}[t]{0.32\textwidth}
        \centering
        \includegraphics[
            width=\linewidth
        ]{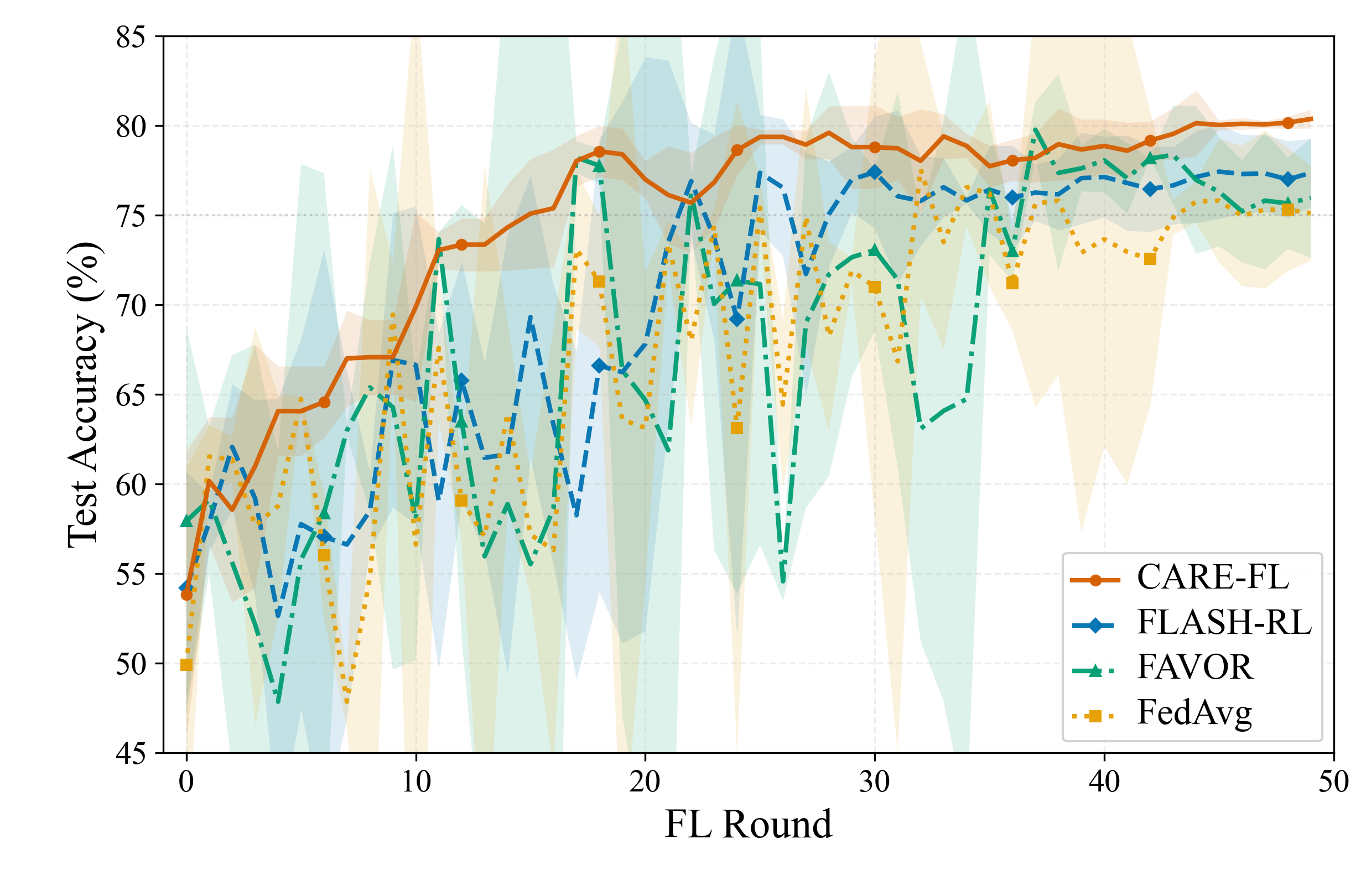}
        \caption{Test accuracy convergence.}
        \label{fig:accuracy_constraint15}
    \end{subfigure}
    \hfill
    % ---------------------------------------------------------
    % (b) Final-stage accuracy
    % ---------------------------------------------------------
    \begin{subfigure}[t]{0.32\textwidth}
        \centering
        \includegraphics[
            width=\linewidth
        ]{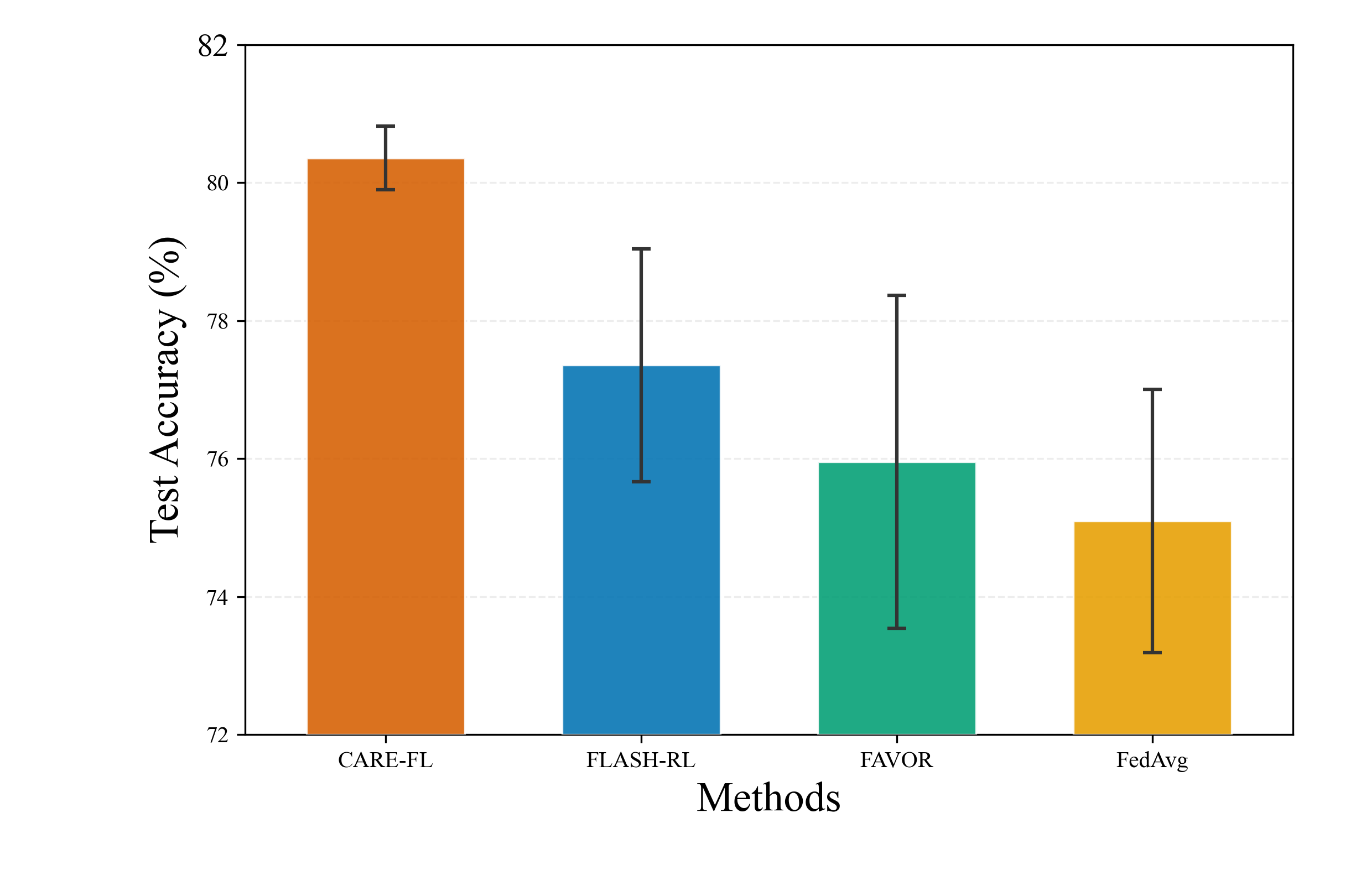}
        \caption{Final-stage accuracy.}
        \label{fig:finalstage_constraint15}
    \end{subfigure}
    \hfill
    % ---------------------------------------------------------
    % (c) Resource usage validation
    % ---------------------------------------------------------
    \begin{subfigure}[t]{0.32\textwidth}
        \centering
        \includegraphics[
            width=\linewidth
        ]{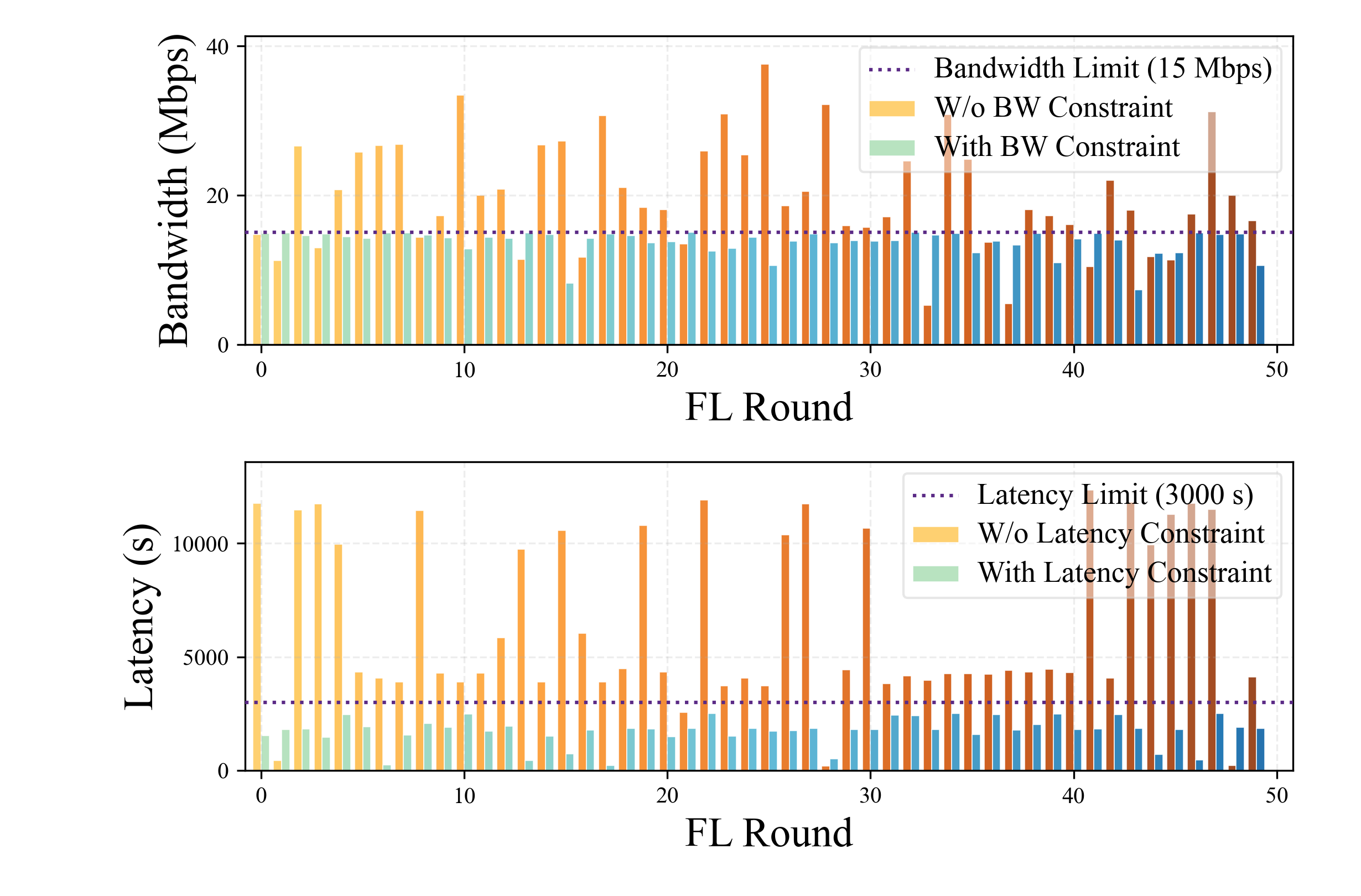}
        \caption{Resource usage validation.}
        \label{fig:resource_constraint15}
    \end{subfigure}

    \caption{Performance comparison and constraint validation under
    moderate resource constraints.}
    \label{fig:combined_constraint15}
\end{figure*}

%figures/0113combined_BW15_Latency_thinConstraint.png
%figures/0113comparison_plot_constraint15_4curves.png

\begin{figure*}[!t]
    \centering

    % ---------------------------------------------------------
    % (a) Test accuracy convergence
    % ---------------------------------------------------------
    \begin{subfigure}[t]{0.32\textwidth}
        \centering
        \includegraphics[
            width=\linewidth
        ]{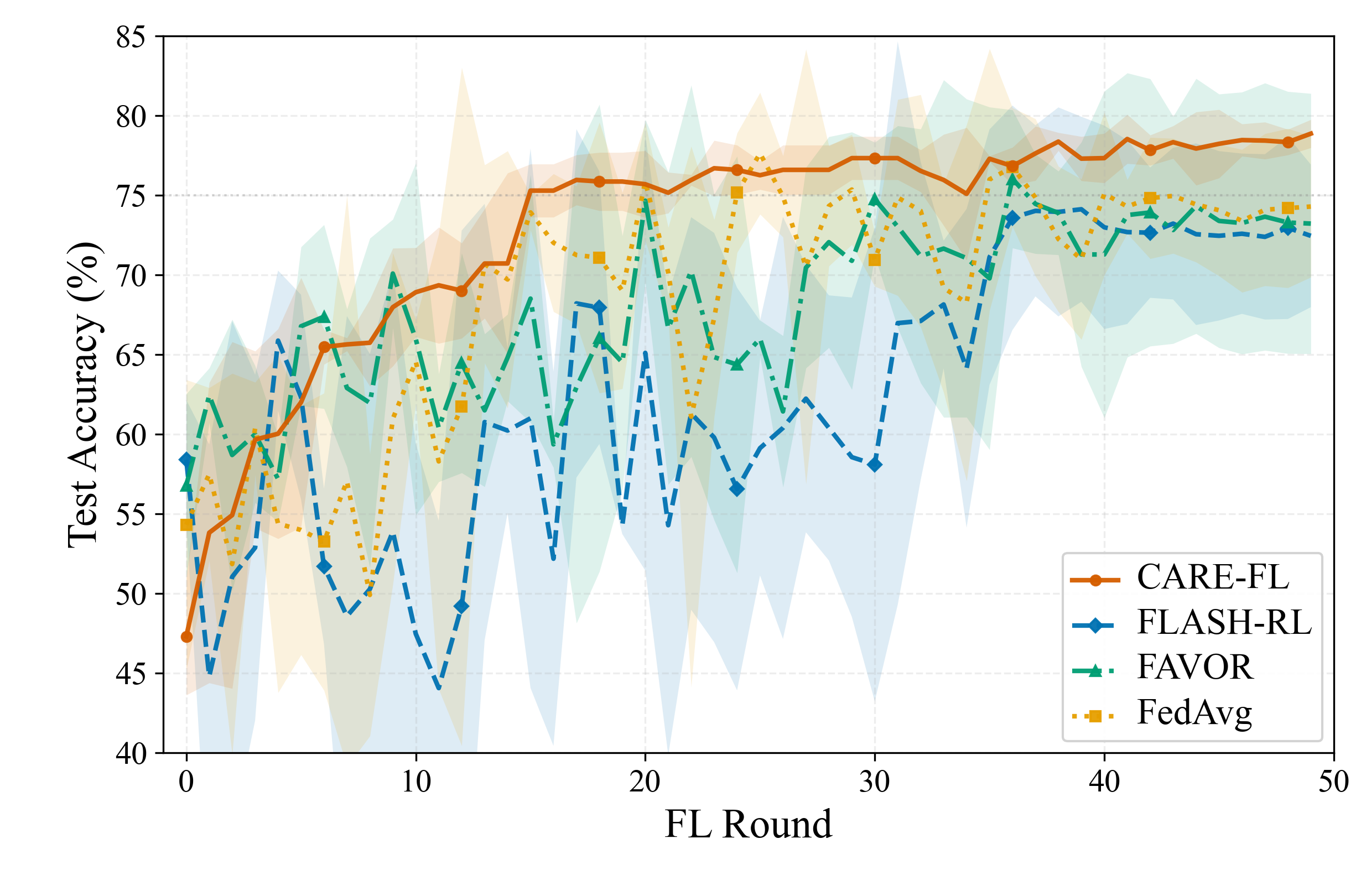}
        \caption{Test accuracy convergence.}
        \label{fig:accuracy_constraint15}
    \end{subfigure}
    \hfill
    % ---------------------------------------------------------
    % (b) Final-stage accuracy
    % ---------------------------------------------------------
    \begin{subfigure}[t]{0.32\textwidth}
        \centering
        \includegraphics[
            width=\linewidth
        ]{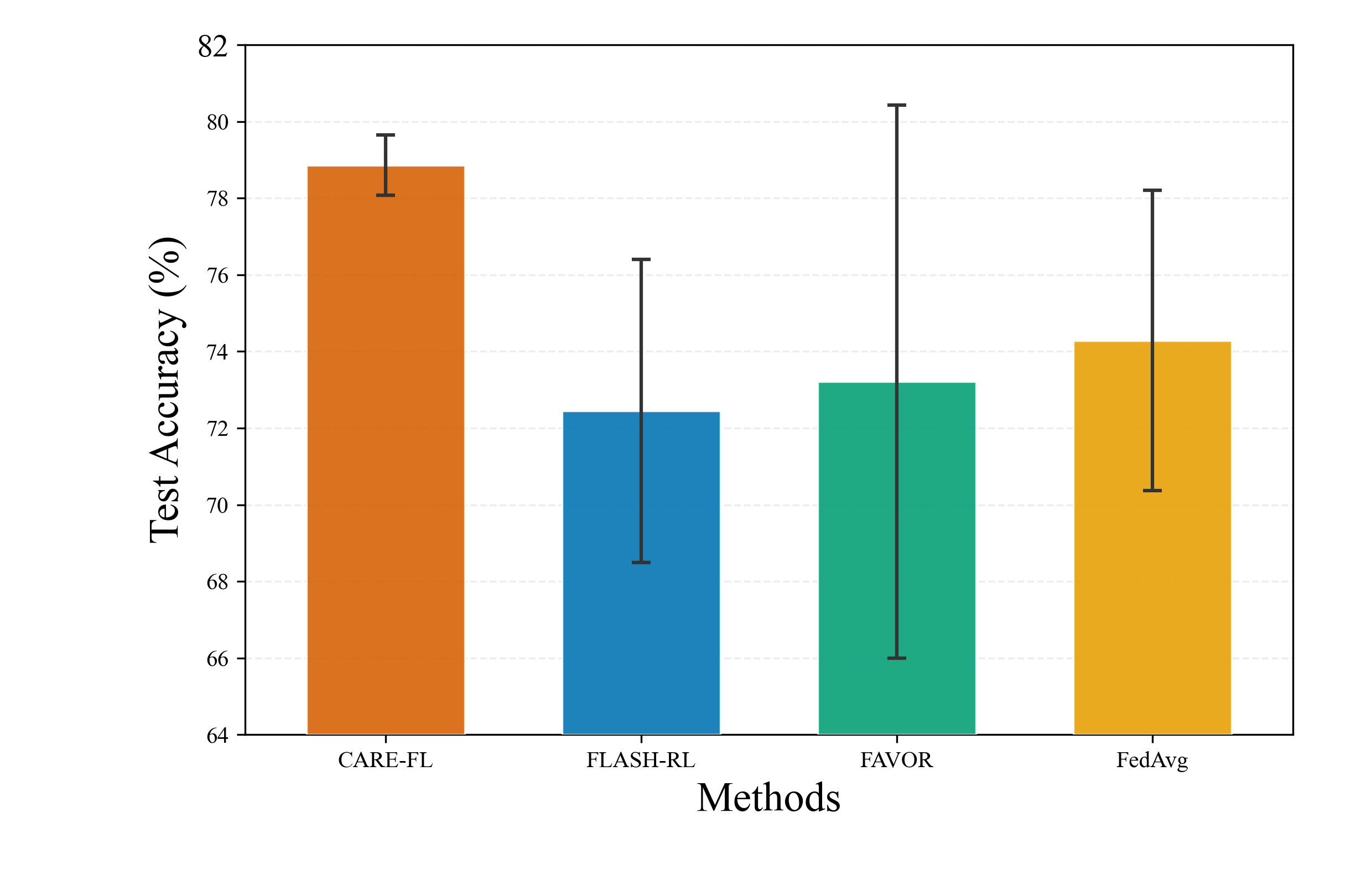}
        \caption{Final-stage accuracy.}
        \label{fig:finalstage_constraint15}
    \end{subfigure}
    \hfill
    % ---------------------------------------------------------
    % (c) Resource usage validation
    % ---------------------------------------------------------
    \begin{subfigure}[t]{0.32\textwidth}
        \centering
        \includegraphics[
            width=\linewidth
        ]{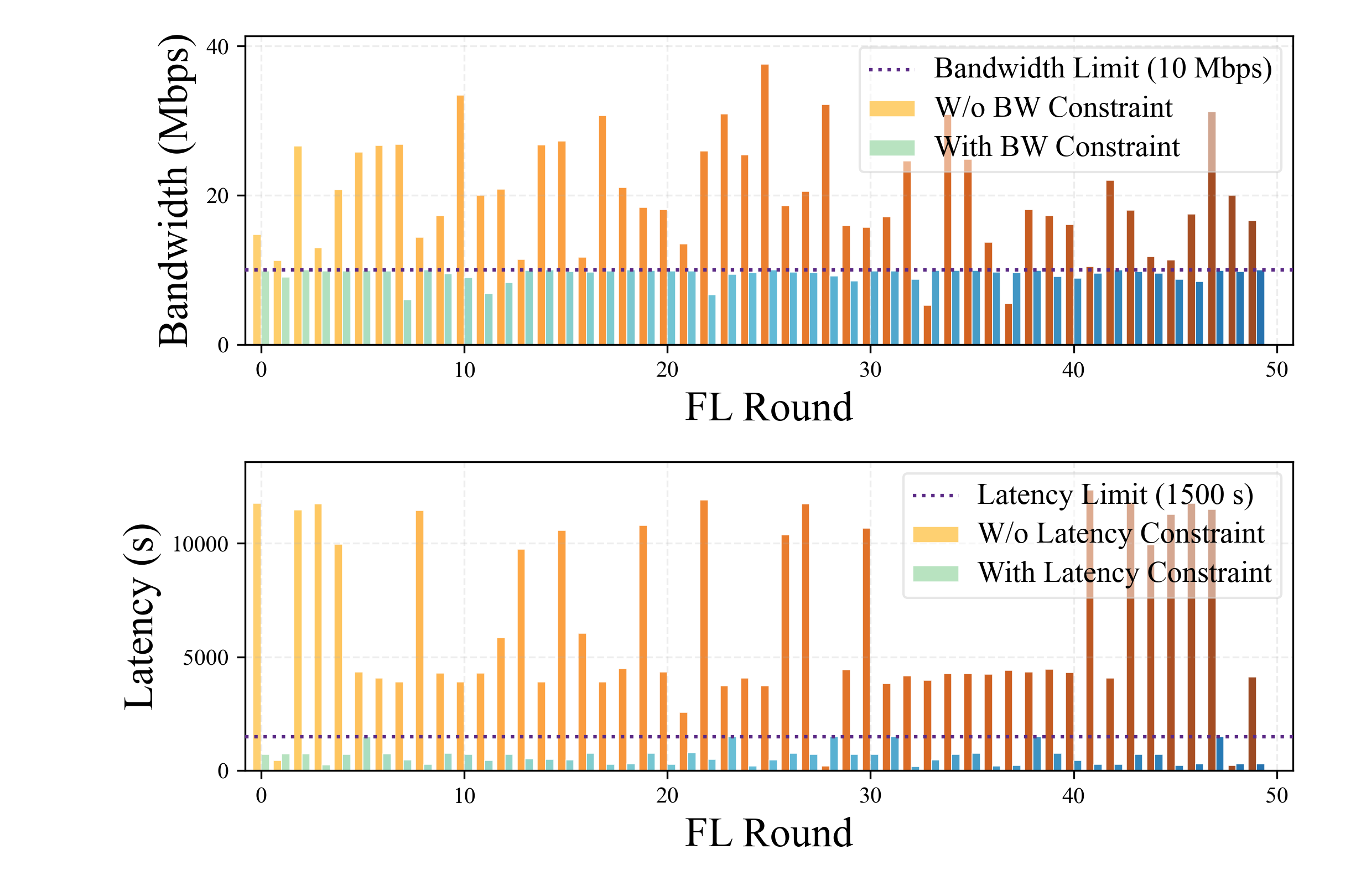}
        \caption{Resource usage validation.}
        \label{fig:resource_constraint15}
    \end{subfigure}

    \caption{Performance comparison and constraint validation under
    stringent resource constraints.}
    \label{fig:combined_constraint15}
\end{figure*}

\subsection{Resource-Constrained Accuracy and Feasibility}

To assess practical deployment, we evaluate whether CARE-FL maintains its performance advantage under wireless resource constraints, examining both convergence and selection feasibility. Figs. 4 and 5 report results under moderate constraints of 15 Mbps and 3000 s and stringent constraints of 10 Mbps and 1500 s, respectively.

Under moderate constraints, Figs. 4(a) and 4(b) show that CARE-FL maintains the strongest late-round accuracy among the compared methods and converges to approximately 80\%.  Tightening the limits to 10 Mbps and 1500 s reduces overall attainable accuracy, but Figs. 5(a) and 5(b) show that CARE-FL still achieves the highest late-round performance, finishing near 79\%. Figs. 4(c) and 5(c) further show that constraint-aware execution maintains bandwidth and latency usage within the prescribed limits, whereas unconstrained selection repeatedly violates them. Thus, CARE-FL preserves a clear convergence advantage while maintaining resource constraints.

These results reflect CARE-FL's constraint-aware selection design. Unlike FLASH-RL, which combines training utility and system cost through reward weights requiring careful balancing, CARE-FL focuses on learning training utilities while explicitly enforcing bandwidth and latency constraints during selection. Tighter constraints shrink the feasible action space, potentially excluding high-utility clients or those with complementary data and modalities. The resulting accuracy reduction illustrates the expected accuracy-resource trade-off. Nevertheless, CARE-FL prioritizes high-utility clients among feasible candidates, maintaining comparatively strong convergence while satisfying resource limits.
%Rather than incorporating system cost into the learned utility through fixed reward-weighting coefficients, \changes{as in FLASH-RL, which requires appropriate balancing of the relative weights of training utility and system cost}, CARE-FL uses the learned training utilities to guide participation while explicitly restricting action selection to subsets satisfying the bandwidth and latency requirements. As the constraints become tighter, the feasible participation space becomes smaller, potentially excluding clients with high estimated training utility or complementary data and modality characteristics. The resulting reduction in accuracy from the moderate to the stringent setting therefore reflects an expected accuracy-resource trade-off. Nevertheless, CARE-FL continues to prioritize high-utility clients within the feasible action space, allowing it to retain comparatively strong convergence while satisfying the imposed resource limits.

\begin{figure*}[t]
    \centering
    % Left Subfigure
    \begin{subfigure}{0.48\textwidth}
        \centering
        % Set a fixed height for both (e.g., 6.8cm)
        \includegraphics[height=5.8cm, width=\linewidth, keepaspectratio=false]{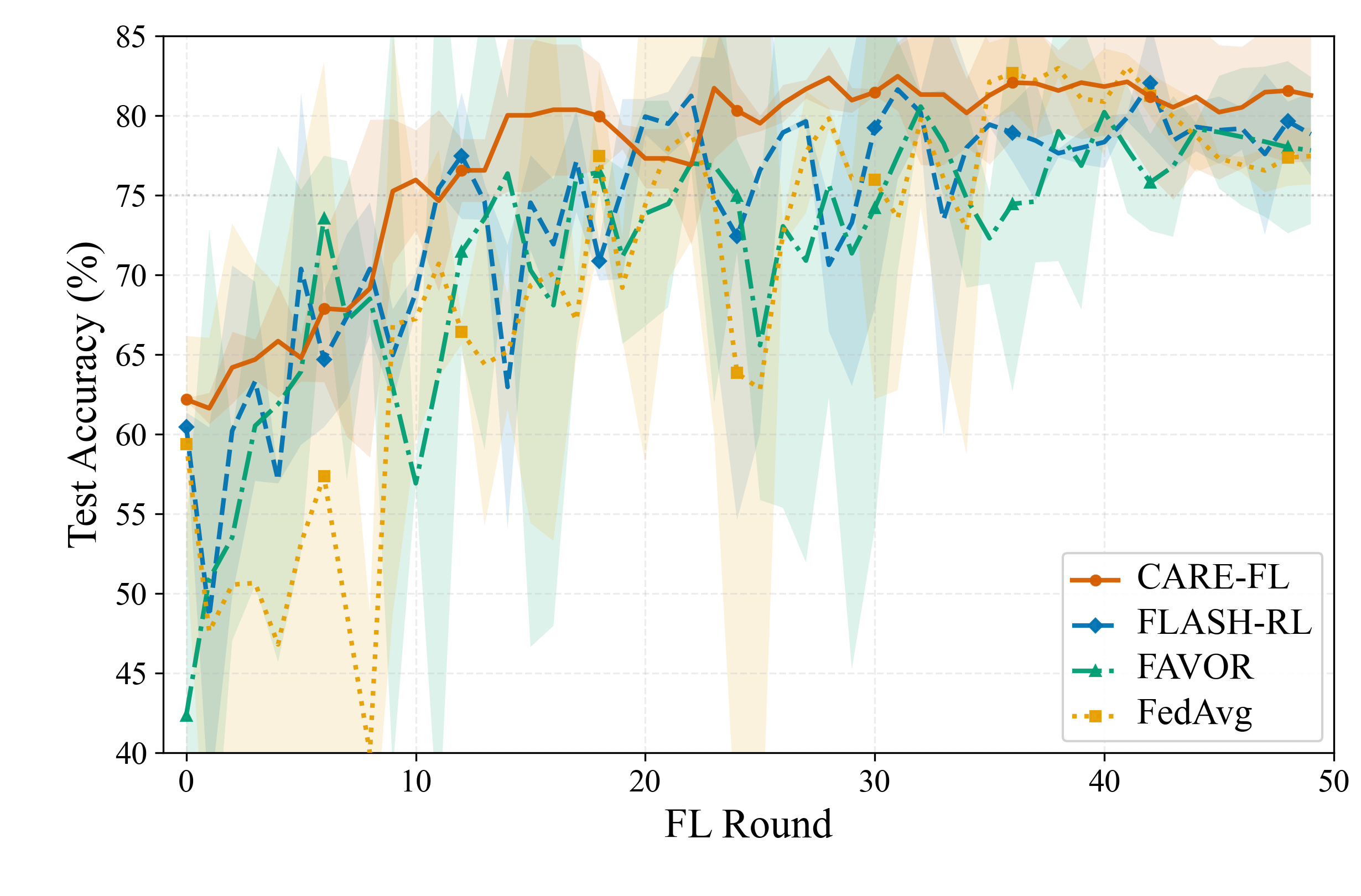}
        \caption{LoRA rank $r=32$ applied to all LLM layers.}
        \label{fig:accuracy_lora_all}
    \end{subfigure}
    \hfill
    % Right Subfigure
    \begin{subfigure}{0.48\textwidth}
        \centering
        % Use the EXACT same height as the left one
        \includegraphics[height=5.8cm, width=\linewidth, keepaspectratio=false]{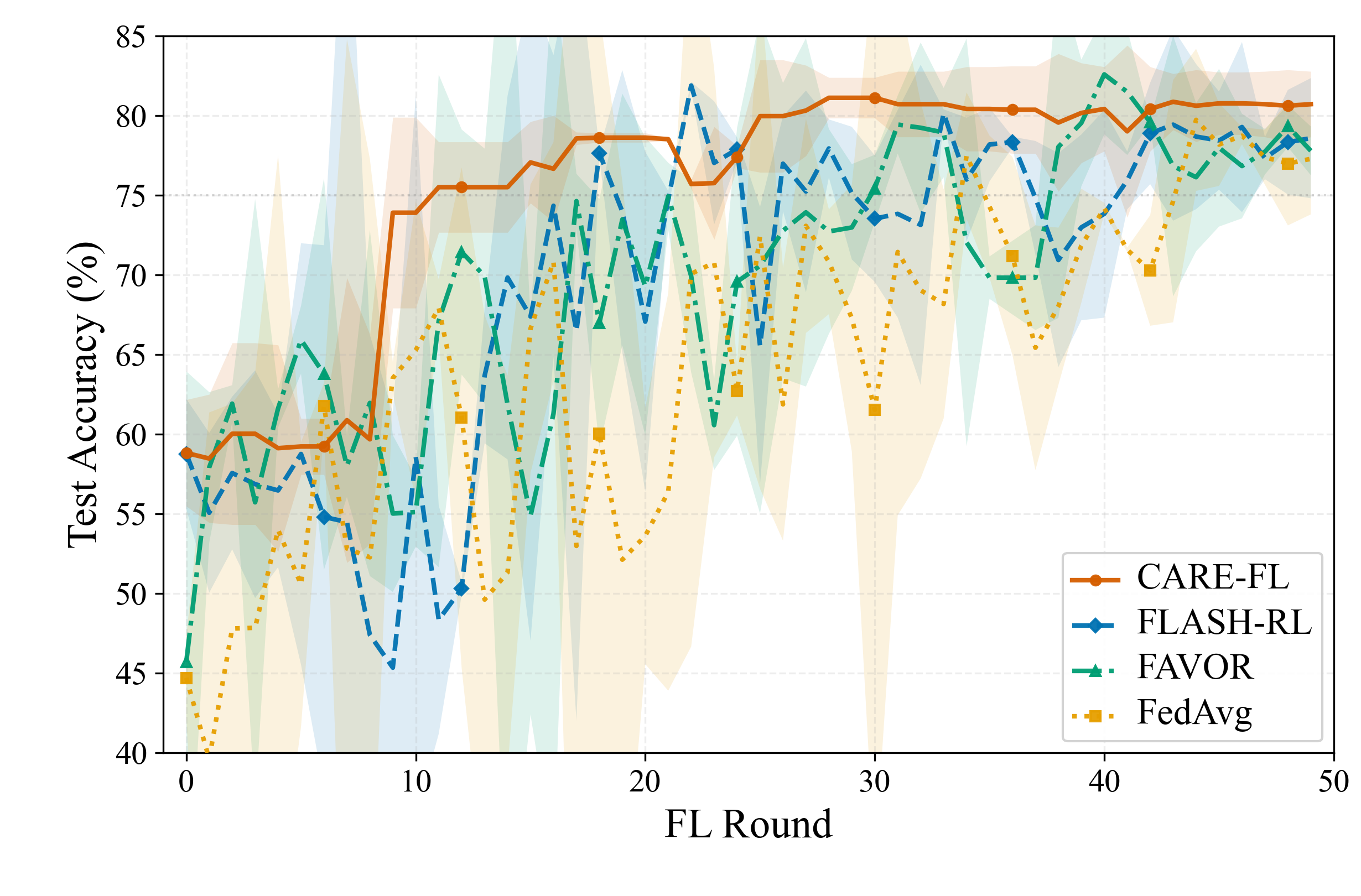}
        \caption{LoRA rank $r=32$ applied to the last 20 LLM layers.}
        \label{fig:accuracy_lora_last20}
    \end{subfigure}
    
   \caption{{Performance comparison under different LoRA configurations.}}
    \label{fig:combined_lora}
\end{figure*}
%figures/0620_ConsBW10_fullLLM_32_acc_curve.png
%figures/0622_ConsBW10_last20LLM_32_acc_curve.png

\begin{figure*}[t]
    \centering

    % =========================================================
    % Left: Independent Fig. 7
    % =========================================================
    \begin{minipage}[t]{0.32\textwidth}
        \centering

        \includegraphics[width=\linewidth]
        {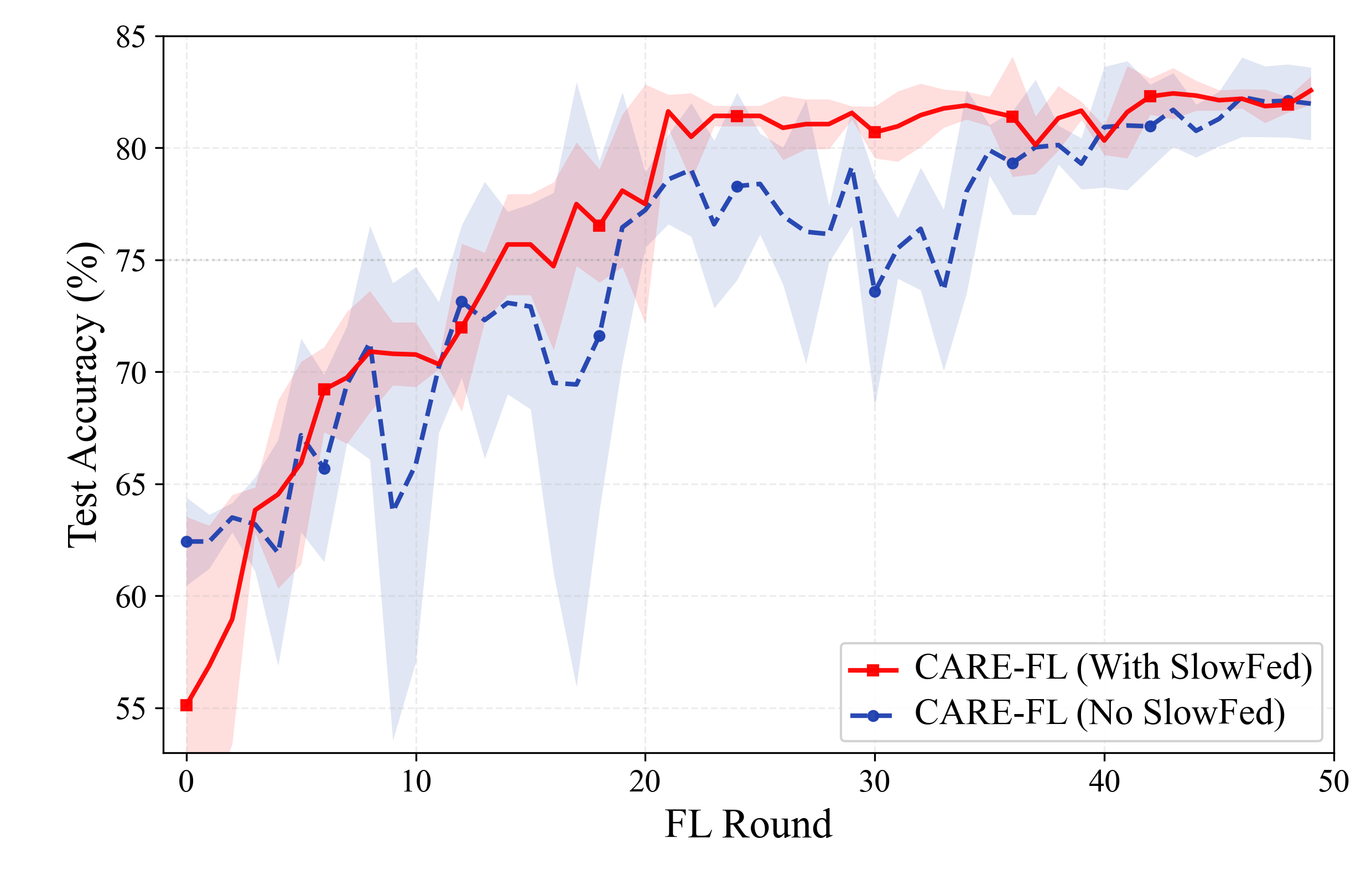}

        \captionof{figure}{
            Ablation study of adaptive aggregation.
            Test accuracy convergence with and without
            adaptive aggregation.
        }
        \label{fig:slowfed_ablation}
    \end{minipage}
    \hfill
    % =========================================================
    % Right: Independent Fig. 8 with two subfigures
    % =========================================================
    \begin{minipage}[t]{0.65\textwidth}
        \centering

        % Fig. 8(a)
        \begin{subfigure}[t]{0.48\linewidth}
            \centering
            \includegraphics[width=\linewidth]
            {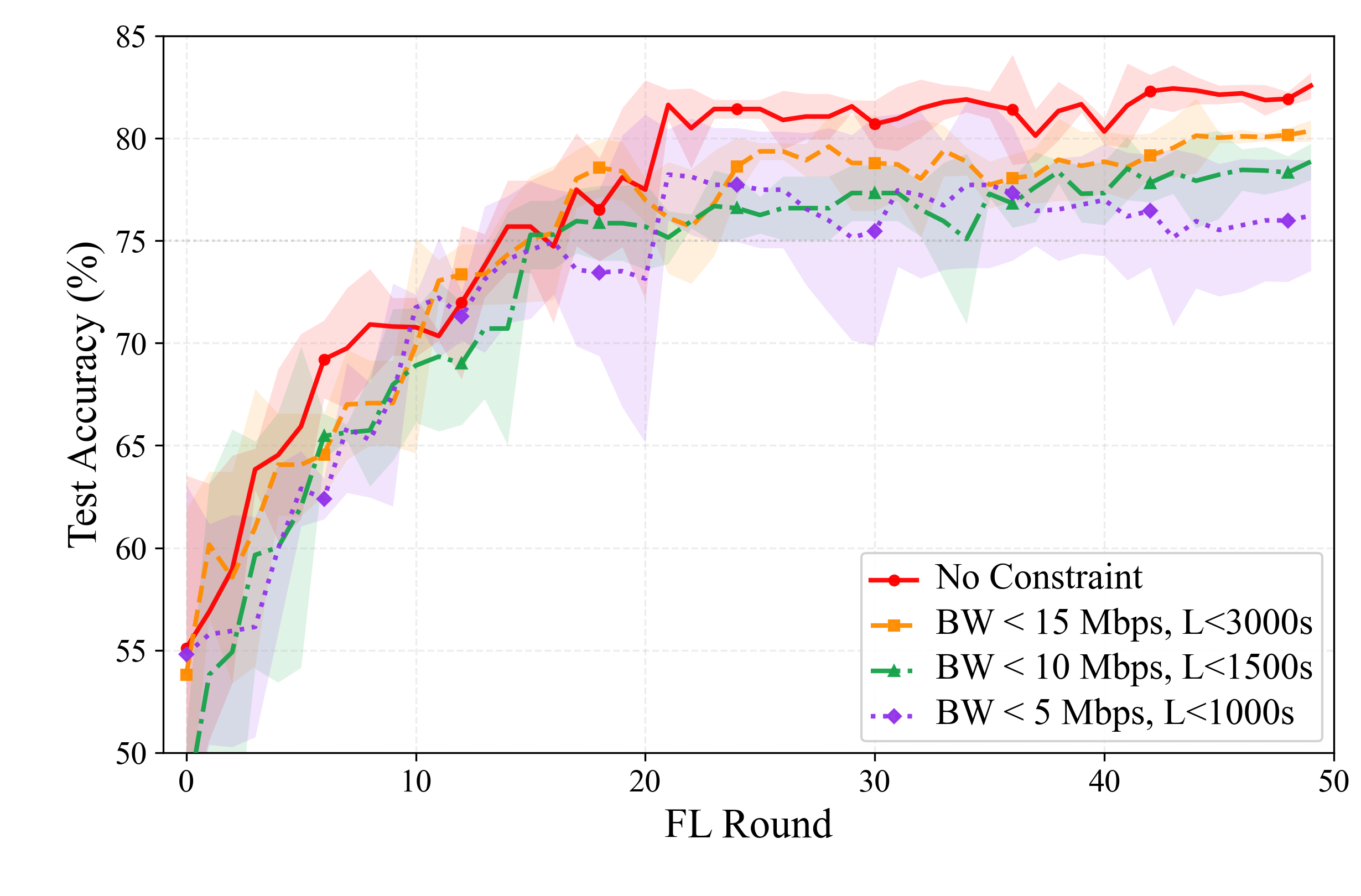}

            \caption{Test accuracy convergence.}
            \label{fig:resource_levels_convergence}
        \end{subfigure}
        \hfill
        % Fig. 8(b)
        \begin{subfigure}[t]{0.48\linewidth}
            \centering
            \includegraphics[width=\linewidth]
            {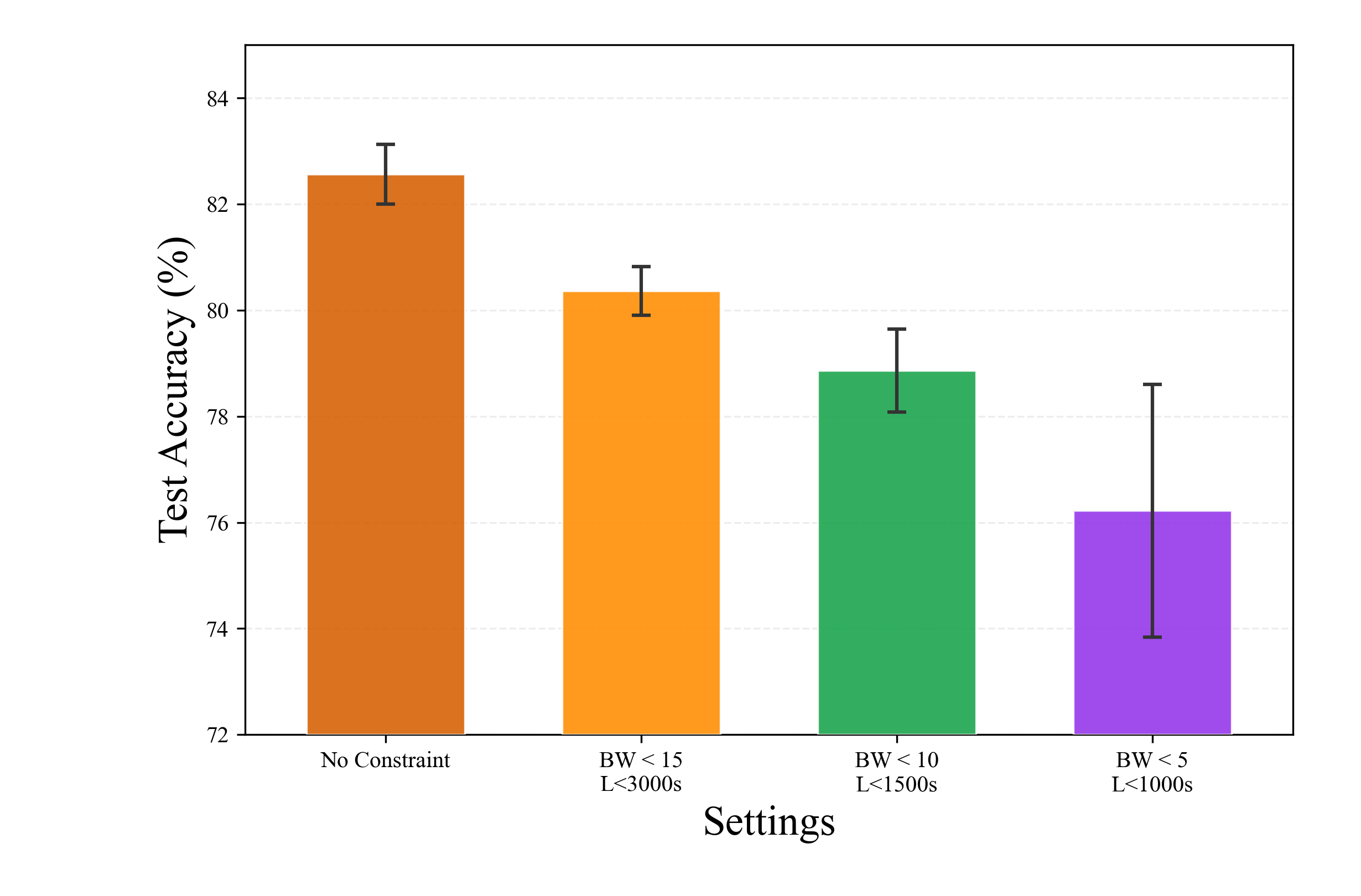}

            \caption{Final-stage accuracy.}
            \label{fig:resource_levels_final}
        \end{subfigure}

        \captionof{figure}{
            Performance under different resource constraint levels.
        }
        \label{fig:resource_levels}
    \end{minipage}

\end{figure*}

%figures/0112comparison_plot_constraint10_4curves.png
%figures/0112combined_BW_Latency_thinConstraint_BW10.png

\subsection{Ablation and Configuration Robustness}

We further assess CARE-FL's robustness across trainable-parameter configurations, the contribution of SlowFed, and different resource constraints. Specifically, Fig. 6 evaluates two additional LoRA rank-\(32\) configurations, Fig. 7 evaluates the effect of SlowFed, and Fig. 8 compares convergence under increasingly stringent resource limits.

Fig. 6 evaluates two configurations with LoRA rank (r=32), applied to either all LLM layers or only the last 20 layers. Compared with the default configuration of \(r=8\) applied to all LLM layers, these rank and layer-placement choices produce different convergence trajectories, reflecting their influence on fine-tuning dynamics. Nevertheless, CARE-FL consistently reaches high accuracy and maintains competitive late-round performance against the baselines under both configurations. These results demonstrate its effectiveness across different PEFT configurations.

Fig. 7 examines SlowFed's contribution. Without SlowFed, CARE-FL exhibits pronounced intermediate-round accuracy drops, whereas SlowFed smooths convergence and reduces these degradations. Both variants approach similar final accuracy, suggesting that SlowFed primarily improves stability rather than guaranteeing higher final accuracy. This aligns with its design: SlowFed evaluates trial global updates using server-side validation performance and progressively reduces the update magnitude when degradation exceeds a prescribed threshold.

Figs. 8(a) and 8(b) compare CARE-FL under increasingly stringent resource constraints. The unconstrained setting achieves the highest late-round accuracy, followed by moderate (15 Mbps, 3000 s), stringent (10 Mbps, 1500 s), and more restrictive (5 Mbps, 1000 s) settings. This ordering highlights the accuracy-resource trade-off: tighter constraints shrink the feasible action space and may exclude high-utility clients or those with complementary data and modalities, limiting attainable accuracy. Nevertheless, CARE-FL maintains comparatively stable convergence by prioritizing high-utility clients within each feasible action space. Together, Figs. 6–8 demonstrate CARE-FL's effectiveness across PEFT and resource configurations, while SlowFed improves robustness against unstable global updates during online and exploratory client selection.

\section{Conclusion and Future Directions}

This paper presented CARE-FL, a constraint-aware reinforcement learning framework for efficient federated fine-tuning of Multimodal Large Language Models (MLLMs) in dynamic wireless edge environments. CARE-FL formulates adaptive client participation as a constrained sequential decision-making problem and employs client-level Q-value decomposition to enable scalable online learning of evolving client utility. A constraint-aware Double Deep Q-Learning (DDQL) policy uses the learned utilities to adapt client participation under dynamic system conditions while satisfying per-round bandwidth and latency constraints. An adaptive aggregation mechanism further mitigates potentially unstable global updates arising during heterogeneous client updates and exploratory participation decisions.

Experiments on the CrisisMMD disaster-response task demonstrate that CARE-FL achieves stronger and more stable convergence than the compared client-selection baselines across both unconstrained and resource-constrained settings. Under bandwidth and latency constraints, CARE-FL maintains comparatively strong model performance while keeping the selected client subsets within the prescribed resource limits. Additional experiments under different LoRA configurations demonstrate the applicability of the proposed framework across different trainable-parameter settings, while the adaptive aggregation ablation confirms its role in improving convergence stability.

%Several limitations motivate further extensions of the proposed framework. Firstly, under stringent per-round latency and bandwidth requirements, resource-limited clients may frequently become infeasible for participation, potentially reducing participation diversity. Addressing this issue while preserving the hard system requirements may require integrating CARE-FL with complementary mechanisms that reduce client computation or communication latency, thereby expanding the set of feasible participants. Secondly, although CARE-FL performs client selection under hard constraints while modeling time-varying client-side system conditions over training rounds, it does not explicitly exploit temporal variations in client latency by scheduling clients during their more resource-efficient periods. Finally, the current adaptive aggregation mechanism relies on server-side validation data to evaluate trial global updates, which may not be available in fully decentralized-data scenarios. 
Future directions include investigating  complementary computation and communication optimization strategies, opportunistic client scheduling and temporal resource prediction, and validation-free stability control to extend CARE-FL to more resource-constrained deployment scenarios.

%% ===========================
\FloatBarrier
\bibliographystyle{IEEEtran}
\bibliography{Reference}

%
%\begin{IEEEbiography}[...]{Huawei~Huang}
%...
%\end{IEEEbiography}
%

\end{document}